\documentclass[times,twocolumn]{aastex62}

\hypersetup{colorlinks=true,citecolor=black,linkcolor=black,urlcolor=black,filecolor=black}

\usepackage{commath}
\usepackage{bm}
\usepackage{mathtools}
\usepackage{amsmath}
\usepackage{mathrsfs}

\newcommand\eq{equation}

\usepackage{acronym}
\newacro{MBH}{massive black hole}
\newacroplural{MBH}[MBHs]{massive black holes}
\newcommand{\MBH}{\ac{MBH}}

\newacro{DF}{distribution function}
\newcommand{\DF}{\ac{DF}}
\newacro{PDF}{probability distribution function}
\newacroplural{PDF}[PDFs]{probability distribution functions}
\newcommand{\PDF}{\ac{PDF}}
\newcommand{\PDFs}{\acp{PDF}}
\newacro{MW}{Milky Way}

\newacro{VRR}{vector resonant relaxation}
\newcommand{\VRR}{\ac{VRR}}
\newacro{SRR}{scalar resonant relaxation}
\newcommand{\SRR}{\ac{SRR}}
\newacro{NSC}{nuclear stellar cluster}
\newacroplural{NSC}[NSCs]{nuclear stellar clusters}
\newcommand{\NSC}{\ac{NSC}}

\newacro{HMF}{Hamiltonian Mean Field}
\newcommand{\HMF}{\ac{HMF}}

\newcommand{\rd}{\mathrm{d}}
\newcommand{\re}{\mathrm{e}}
\newcommand{\ri}{\mathrm{i}}
\newcommand{\rc}{\mathrm{c}}
\newcommand{\rt}{\mathrm{t}}

\newcommand{\mJ}{\mathcal{J}}
\newcommand{\aout}{a_{\mathrm{out}}}
\newcommand{\ain}{a_{\mathrm{in}}}
\newcommand{\bL}{\widehat{\mathbf{L}}}
\newcommand{\phip}{\phi^{\prime}}
\newcommand{\ein}{e_{\mathrm{in}}}
\newcommand{\eout}{e_{\mathrm{out}}}
\newcommand{\mBH}{M_{\bullet}}
\newcommand{\bK}{\mathbf{K}}
\newcommand{\bX}{\mathbf{X}}
\newcommand{\vphi}{\varphi}
\newcommand{\deltaD}{\delta_{\mathrm{D}}}
\newcommand{\bLp}{\widehat{\mathbf{L}}^{\prime}}
\newcommand{\bKp}{\mathbf{K}^{\prime}}
\newcommand{\bKpp}{\mathbf{K}^{\prime\prime}}
\newcommand{\ellp}{\ell^{\prime}}
\newcommand{\mpp}{m^{\prime}}
\newcommand{\tp}{t^{\prime}}
\newcommand{\avEA}[1]{\left\langle #1 \right\rangle}
\newcommand{\savEA}[1]{\langle #1 \rangle}

\newcommand{\avT}[1]{\left\langle #1 \right\rangle_{T}}
\newcommand{\savT}[1]{\langle #1 \rangle_{T}}
\newcommand{\navT}[1]{\big\langle #1 \big\rangle_{T}}
\newcommand{\bavT}[1]{\bigg\langle #1 \bigg\rangle_{T}}
\newcommand{\avW}[1]{\left\langle #1 \right\rangle_{W}}
\newcommand{\avWt}[1]{\left\langle #1 \right\rangle_{W_{\mathrm{t}}}}
\newcommand{\savWt}[1]{\langle #1 \rangle_{W_{\mathrm{t}}}}

\newcommand{\avLt}[1]{\left\langle #1 \right\rangle_{\bLt}}
\newcommand{\savLt}[1]{\langle #1 \rangle_{\bLt}}
\newcommand{\avc}[1]{\left\langle #1 \right\rangle_{\mathrm{c}}}

\newcommand{\bLt}{\widehat{\mathbf{L}}_{\mathrm{t}}}
\newcommand{\bKt}{\mathbf{K}_{\mathrm{t}}}
\newcommand{\vphit}{\varphi^{\mathrm{t}}}
\newcommand{\Qt}{Q^{\mathrm{t}}}

\newcommand{\veps}{\varepsilon}

\newcommand{\Tc}{T_{\mathrm{c}}}
\newcommand{\erf}{\text{erf}}

\newcommand{\mO}{\mathcal{O}}
\newcommand{\mmin}{m_{\mathrm{min}}}

\newcommand{\amin}{a_{\mathrm{min}}}
\newcommand{\amax}{a_{\mathrm{max}}}
\newcommand{\emin}{e_{\mathrm{min}}}
\newcommand{\emax}{e_{\mathrm{max}}}
\newcommand{\ophi}{\overline{\vphi}}
\newcommand{\rr}{\mathrm{r}}
\newcommand{\dF}{\mathbb{F}}
\newcommand{\vphip}{\vphi^{\prime}}
\newcommand{\bvphi}{\bm{\vphi}}
\newcommand{\bvphip}{\bm{\vphi}^{\prime}}
\newcommand{\Ep}{E^{\prime}}

\newcommand{\bT}{\bm{\theta}}
\newcommand{\bTp}{\bm{\theta}^{\prime}}

\newcommand{\num}{\mathrm{num.}}
\newcommand{\pred}{\mathrm{pred.}}
\newcommand{\br}{\mathbf{r}}
\newcommand{\rp}{r_{\mathrm{p}}}
\newcommand{\ra}{r_{\mathrm{a}}}
\newcommand{\bmeta}{\bm{\eta}}
\newcommand{\rh}{r_{\mathrm{h}}}
\newcommand{\pc}{\mathrm{pc}}
\newcommand{\mpsq}{m^{\prime 2}}
\newcommand{\avmsq}{\langle m^{2} \rangle}
\newcommand{\app}{a^{\prime}}
\newcommand{\epp}{e^{\prime}}
\newcommand{\eppsq}{e^{\prime 2}}

\shorttitle{Vector resonant relaxation}
\shortauthors{Fouvry, Bar-Or, Chavanis}

\begin{document}

\title{Vector Resonant Relaxation of Stars around a Massive Black Hole}

\author{Jean-Baptiste Fouvry}
\altaffiliation{Hubble Fellow}
\affiliation{Institute for Advanced Study, Princeton, NJ, 08540, USA}
\author{Ben Bar-Or}
\affiliation{Institute for Advanced Study, Princeton, NJ, 08540, USA}
\author{Pierre-Henri Chavanis}
\affiliation{Laboratoire de Physique Th\'eorique (IRSAMC), CNRS and UPS, Univ. de Toulouse, France, F-31062 Toulouse, France}


\begin{abstract}
  In the vicinity of a massive black hole, stars move on precessing Keplerian
  orbits. The mutual stochastic gravitational torques between the stellar
  orbits drive a rapid reorientation of their orbital planes, through a process
  called vector resonant relaxation. We derive, from first principles, the
  correlation of the potential fluctuations in such a system, and the
  statistical properties of random walks undergone by the 
  stellar orbital orientations.
  We compare this new analytical approach
  with effective $N$-body simulations.
  We also provide a simple scheme to
  generate the random walk of a test star's orbital orientation
  using a stochastic equation of motion.
  We finally present quantitative estimations of this process for a nuclear stellar cluster
  such as the one of the Milky Way.
\end{abstract}

\section{Introduction}
\label{sec:Introduction}

Most nearby galaxies possess a \MBH\ in their center, surrounded by a
\NSC~\citep{Genzel2010,KormendyHo2013,Graham2016}. The dynamical evolution of
the stellar cluster comprises numerous processes acting on different
timescales~\citep{RauchTremaine1996,HopmanAlexander2006,Alexander2017}~\citep[see
also Fig.~{1} in][]{Kocsis2011}. Since the gravitational potential is dominated
by the central \MBH, stars move on nearly Keplerian orbits. The deviations from
a Keplerian potential due to the stellar potential and the relativistic
corrections, cause the Keplerian ellipses to precess in their orbital plane.
Subsequently, through the non-spherical components of the potential
fluctuations, the orbital orientation of the stars get reshuffled, without changing the
magnitude of their angular momentum nor their Keplerian energy, through a
process called \VRR~\citep[][and references therein]{Kocsis2015}, which is the
focus on this work.  Resonant torques' coupling between the precessing stars
then lead to a diffusion of the stars' angular momentum magnitude, a process
called \SRR~\citep[][and references therein]{BarOr2018}. Finally, on longer
timescales, close encounters between stars lead to the relaxation of the stars'
Keplerian energy and angular momentum~\citep{Bahcall+1976,
  Bahcall+1977, Lightman+1977, Cohn+1978, Shapiro+1978}.

\VRR\ can be a driving force behind several dynamical phenomena in galactic
centers, including the warping of accretion~\citep{Bregman+2009, Bregman+2012}
and stellar~\citep{Kocsis2011} disks, as well as a catalyzer of binaries
mergers~\citep{Hamers2018}.
Since its first presentation by~\cite{RauchTremaine1996}, \VRR\ was studied
numerically both with full~\citep{Eilon+2009} and effective
orbit-averaged~\citep{Kocsis2015} $N$-body simulations.
More recently,~\cite{Roupas2017,Takacs2018, Szolgyen2018}
studied the thermodynamical equilibria of \VRR\@,
and~\cite{FouvryBarOrChavanis2018} its axisymmetric limit.

In the present paper, building upon these works, we set out to offer a detailed
characterization of the \VRR\ process in the limit of an isotropic distribution
of stars. To do so, in Section~\ref{sec:Model}, we present the fundamental
equations of \VRR\@. In Section~\ref{sec:CorrelNoise}, we characterize the
properties of the potential fluctuations in the system, as inferred from
estimates of the correlation function at the initial time. This will allow us
then to describe in Section~\ref{sec:TestParticle} the random walk of a test
particle's orientation, and develop an effective stochastic equation of motion
which can efficiently mimic these random motions. Detailed comparisons of
these results with effective numerical simulations are presented throughout
these sections. In Section~\ref{sec:SelfConsistency}, we detail the important
self-consistency existing between the potential fluctuations and the properties
of the orientations' random walks.  Finally, in Section~\ref{sec:PowerLaw}, we
use this new formalism to present the timescales associated with \VRR\ in a
nuclear stellar cluster similar to the Milky Way's. We conclude in
Section~\ref{sec:Conclusion}.

\section{Model}
\label{sec:Model}

\begin{figure}
\begin{center}
\includegraphics[width=0.45\textwidth]{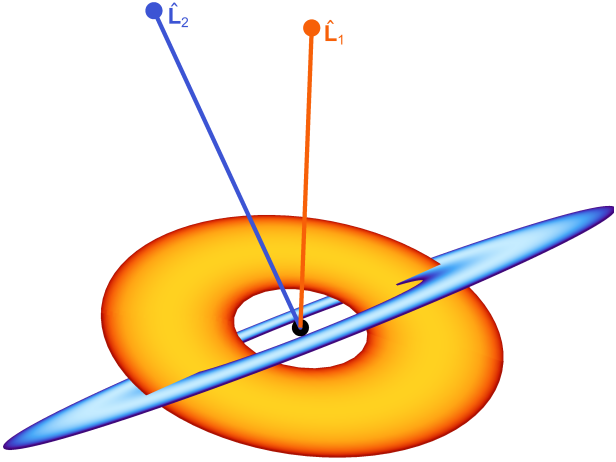}
\caption{
Illustration of the orbit averaged interaction between two stars orbiting a central supermassive object.
Following the average over the fast Keplerian motion and the in-plane
precession, stars are replaced by annuli, where darker colors indicate a higher
surface density (not to scale).
The interaction between two annuli then depends on each star's conserved parameters ${ \bK = (m , a ,e) }$, as well as on their respective orbital orientations given by the normal vectors $\bL_{1}$ and $\bL_{2}$.
\label{fig:IllustrationInteraction}}
\end{center}
\end{figure}

We consider a set of $N$ stars orbiting a \MBH\ of mass $\mBH$. On timescales
longer than the in-plane precession but shorter than the time to change the
orbital eccentricity (by \SRR\@) and the time to change the semi-major axis (by
two-body relaxation) the mutual interactions between two stars can be orbit-averaged
over their respective mean anomalies and in-plane precession angles.
As a result, each star can be replaced by a disk of mass $m$ extending between
${ \rp \!=\! (1-e) a }$ and ${ \ra \!=\! (1+e) a }$ with surface density
${ \Sigma(r) \!=\! {[2\pi^2 a \sqrt{r \!-\! \rp}\sqrt{\ra \!-\! r}]}^{-1} }$,
where the semi-major axis $a$ and eccentricity $e$ can be assumed to be
constant in time (see an illustration in Fig.~\ref{fig:IllustrationInteraction}).
Following this double orbit-average, one can associate to each star a set of
conserved quantities ${ \bK = (m, a, e) }$ and a time-dependent normal vector
${ \bL }$, with $\bm{L} = L \bL$ the orbital angular momentum.  We introduce
the spherical coordinates as ${ (\theta , \phi) }$, so that
${ \bL = (\sqrt{1-u^2}\cos\phi, \sqrt{1-u^2}\sin\phi, u) }$, with
${ u = \cos (\theta) }$.  Studying \VRR\ amounts then to studying the long-term
dynamics of each star's normal vector $\bL$.

Following~\cite{Kocsis2015}, the effective single particle Hamiltonian of \VRR\
reads
\begin{align}
H & \, = \sum_{i=1}^{N} \bigg\langle{- \frac{G m m_i}{|\br (t) - \br_{i} (\tp)|} \bigg\rangle}_{t , \tp}
\nonumber
\\
&\, = -  L(\bK)  \sum_{\mathclap{\substack{\ell \geq 2 \\
   \mathrm{even}}}} \sum_{m=-\ell}^{\ell} M_{\ell m} ( \bK , t) \, Y_{\ell m} (\bL) .
\label{Hamiltonian_VRR}
\end{align}
with ${ \br (t) }$, ${ \br_{i} (t) }$ the positions of the test star and the star $i$,
as they move along their (in-plane) precessing Keplerian orbits,
${ \langle \, \cdot \, \rangle_{t , \tp} }$ the double orbit-average over these
motions,
and $\bK$ and $\bK_{i}$ their respective conserved parameters.
In the second line of Eq.~\eqref{Hamiltonian_VRR}, we introduced the magnetizations 
\begin{equation}
M_{\ell m} (\bK, t) =\sum_{i=1}^N \mJ_{\ell} \big[ \bK , \bK_{i} \big] \, Y_{\ell m} (\bL_{i} (t)),
\label{def_Mlm}
\end{equation}
where the coupling coefficients ${ \mJ_{\ell} [ \bK , \bK_{j} ] }$ are
defined in Eq.~\eqref{def_mJ}, and we used real spherical harmonics
${ Y_{\ell m} (\bL) }$ (defined in Eq.~\eqref{def_ReYlm}).
It is also important to note that only even harmonics with ${ \ell \geq 2 }$
contribute the particles' dynamics, in virtue of the symmetries of the
interaction.

Hamilton's equations of motion read
\begin{equation}
  \label{eq:EOM}
  \dpd{\phi}{t} = \dpd{H}{L_z}; \;\;\; \dpd{L_z}{t} = -\dpd{H}{\phi},
\end{equation}
where ${ L_z = L u }$ is an action and $\phi$ its conjugated angle. The
evolution of the angular momentum of a single test particle is given by
\begin{equation}
\frac{\rd \bm{L}}{\rd t} = \dpd{}{\bm{L}} \times (H \bm{L}) = L \sum_{\ell , m}
M_{\ell m} (\bK, t) \, \bX_{\ell m} (\bL (t)) ,
\label{good_EOM}
\end{equation}
where ${ \bX_{\ell m} (\bL) = \bL \!\times\! \partial Y_{\ell m} (\bL) / \partial \bL
}$ are the real vector spherical harmonics.
Because $L$ is constant, one has ${ \rd \bL /  \rd t \!=\! L^{-1} \rd \bm{L} /  \rd t }$.

Inspired by~\cite{Klimontovich1967}, the state of the system of $N$ stars at time $t$
is fully characterized by the discrete \DF\
\begin{equation}
\vphi (\bL , \bK , t) = \sum_{i} \deltaD (\bL - \bL_{i} (t)) \, \deltaD (\bK - \bK_{i}) ,
\label{def_vphi}
\end{equation}
with${ \deltaD }$ the Dirac delta,
${ \deltaD (\bL \!-\! \bL_{i}) \!=\! \deltaD (u \!-\! u_{i}) \deltaD (\phi
  \!-\! \phi_{i}) }$, and
${ \deltaD (\bK \!-\! \bK_{i}) \!=\! \deltaD (m \!-\! m_{i}) \deltaD (a \!-\!
  a_{i}) \deltaD (e \!-\! e_{i}) }$, with the associated volumes
${ \rd \bL = \rd u \rd \phi }$ and ${ \rd \bK = \rd m \rd a \rd e }$.  The
continuity equation,
${ \partial \vphi / \partial t = - \partial / \partial \bL \!\cdot\! [ \vphi \,
  \partial \bL / \partial t] }$, gives us then
\begin{align}
\frac{\partial \vphi (\bL , \bK , t)}{\partial t} = - & \, \sum_{\ell , m} \!\! \int \!\! \rd \bLp \rd \bKp \, \vphi (\bLp , \bKp , t) \, \mJ_{\ell} \big[ \bK , \bKp \big]
\nonumber
\\
& \, \times Y_{\ell m} (\bLp) \, \bX_{\ell m} (\bL) \cdot \frac{\partial \vphi (\bL , \bK , t)}{\partial \bL} ,
\label{EOM_vphi_I}
\end{align}
where we used the fact that the vector spherical harmonics satisfy ${ \partial / \partial \bL \!\cdot\! \bX (\bL) = 0 }$.
This equation can subsequently be developed in spherical harmonics, by writing
\begin{equation}
\vphi (\bL , \bK , t) = \vphi_{\alpha} (\bK , t) \, Y_{\alpha} (\bL) ,
\label{def_vphiellm}
\end{equation}
where the sum over the index ${ \alpha = (\ell_{\alpha} , m_{\alpha}) }$ is implied, and we introduced
\begin{equation}
\vphi_{\alpha} (\bK , t) = \sum_{i} \deltaD (\bK - \bK_{i}) \, Y_{\alpha} (\bL_{i} (t)) .
\label{def_vphi_alpha}
\end{equation}

When expanded in spherical harmonics, Eq.~\eqref{EOM_vphi_I} becomes
\begin{equation}
\frac{\partial \vphi_{\alpha} (\bK , t)}{\partial t} = - \!\! \int \!\! \rd \bKp \, Q_{\alpha \gamma \delta} (\bK , \bKp) \, \vphi_{\gamma} (\bKp , t) \, \vphi_{\delta} (\bK , t) ,
\label{EOM_vphi_II}
\end{equation}
where the sums over the harmonic indices ${ (\gamma , \delta ) }$ are implied, and we introduced the time-independent coupling tensor ${ Q_{\alpha \gamma \delta} (\bK , \bKp) \!=\! \mJ_{\ell_{\gamma}} [\bK , \bKp] \, E_{\alpha \gamma \delta} }$, with $E_{\alpha \gamma \delta}$ the (real) Elsasser coefficients~\citep{James1973} (see Appendix~\ref{sec:Elsasser} for their properties)
\begin{equation}
E_{\alpha \gamma \delta} = \!\! \int \!\! \rd \bL \, Y_{\alpha} (\bL) \, \bX_{\gamma} (\bL) \cdot \frac{\partial Y_{\delta} (\bL)}{\partial \bL} .
\label{definition_Elsasser_coefficients}
\end{equation}
Equation~\eqref{EOM_vphi_II} is an exact writing of the fundamental evolution equation for \VRR\@.
Its complexity stems in particular from being a quadratic matrix differential equation in the fields ${ \vphi_{\alpha} (\bK , t) }$.

\begin{figure}
\begin{center}
\includegraphics[width=0.35\textwidth]{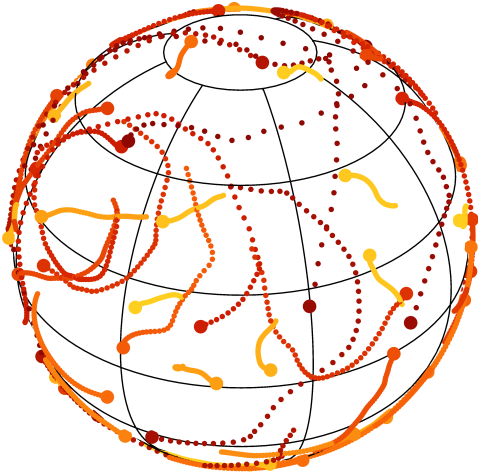}
\caption{
Illustration of the random walk in orientation of a sample of particles from one fiducial simulation.
The orientation of the particles is represented every ${20 h}$, with $h$ the integration timestep (see Appendix~\ref{sec:Nbody}).
Particles are colored according to their semi-major axis (from red for small $a$ to yellow for large $a$).
Particles with larger $a$ see their orientation evolve slower, as a result of the ${ 1/L }$ prefactor in the interaction coefficients of Eq.~\eqref{def_mJ}.
Section~\ref{sec:CorrelNoise} characterizes the properties of the potential fluctutations jointly created by this large collection of particles.
\label{fig:IllustrationSystem}}
\end{center}
\end{figure}

All the upcoming derivations will be illustrated by comparisons with direct
$N$-body simulations.  In Appendix~\ref{sec:Nbody}, we present the fiducial
system considered, as well as the details of our numerical implementation. In
Fig.~\ref{fig:IllustrationSystem}, we illustrate a subset of trajectories from
one such simulation.

\section{The correlation function of the noise}
\label{sec:CorrelNoise}

As a first step towards the characterization of the correlated stochastic
dynamics of one star in that system, we focus our interest on describing the
properties of the density fluctuations generated as a whole by the system's $N$
particles. In particular, we will show how one can use estimates
of the derivatives of the correlation function of the density fluctuations at
the initial time to provide a sensible ansatz (see Eq.~\eqref{Gaussian_Cbath})
for the time dependences of this same correlation function.

The harmonic coefficients ${ \vphi_{\alpha} (\bK , t) }$ (Eq.~\ref{def_vphi_alpha}) describe the full
state of the ${ N \gg 1 }$ particles system at time $t$ and can therefore be treated as
stochastic density fluctuations, assumed to be Gaussian random fields.
Assuming that the system's evolution is stationary in time, the properties of
these fluctuations are captured by the correlation function
\begin{equation}
\label{def_C}
C_{\alpha \beta} (\bK , \bK' , t - \tp) \equiv  \avEA { \vphi_{\alpha} (\bK , t) \, \vphi_{\beta} (\bK' , \tp) } ,
\end{equation}
where ${ \avEA{\cdot} }$ is the ensemble average over realizations (initial conditions
and trajectories of the $N$ particles).

The correlation function is even and generically decreases to zero on a timescale larger than
some coherence time $\Tc$.
As a result, as a first approximation, it is therefore reasonable to replace
${ C_{\alpha \beta} (\bK , \bK' , t - \tp) }$ by a Gaussian function,
tailored to match the function's behavior for ${ t \ll \Tc }$,
see Fig.~\ref{fig:CorrelationOphi} for a justification.

In Appendix~\ref{DerivativesNoise}, we compute the first two derivatives
of the correlation function, and we show in Eqs.~\eqref{res_C0} and~\eqref{res_C2}
that
\begin{equation}
\label{C0_bath}
  C_{\alpha \beta} (\bK , \bK' , 0) = \delta_{\alpha}^{\beta} \, \deltaD (\bK - \bK') \, n (\bK) 
\end{equation}
and
\begin{align}
\label{calc_C2_III_bath}
\frac{\partial^{2}}{\partial t^{2}} C_{\alpha \beta} (\bK , \bK' , t) \bigg|_{t = 0} \!\!\! =
    - \delta_{\alpha}^{\beta} \, A_{\ell_{\alpha}} \deltaD (\bK- \bK') \,
     n (\bK) \, \Gamma^2(\bK),
\end{align}
with the coefficient
\begin{equation}
A_{\ell} = \ell (\ell + 1) .
\label{def_Al}
\end{equation}
In Eq.~\eqref{calc_C2_III_bath}, we introduced ${ n (\bK) }$,
the \DF\ of the stars’ ${ \bK \!=\! (a,e,m) }$ parameters,
that satisfies the normalization convention ${ \!\int\! \rd \bL \rd \bK n(\bK) \!=\! N }$.
We also introduced the decay rate of the correlation function $\Gamma(\bK)$ as
\begin{align}
\label{Gamma_bath}
\Gamma^2(\bK) = \!\! \int \!\! \rd \bK' \, n (\bKp) \, \sum_{\ell} \! B_{\ell} \, \mJ_{\ell}^{2} [\bK , \bKp] ,
\end{align}
with the coefficient
\begin{equation}
\label{def_Bl}
B_{\ell} = \frac{\ell (\ell + 1) (2 \ell + 1)}{8 \pi} .
\end{equation}

Gathering Eqs.~\eqref{C0_bath} and~\eqref{calc_C2_III_bath}, we can approximate the correlation function $C_{\alpha \beta} (\bK ,
\bK' , t)$ by 
\begin{equation}
  \label{eq:C_bath}
  C_{\alpha \beta} (\bK , \bK' , t) = \delta_\alpha^\beta \, \deltaD(\bK - \bKp) \, C_{\ell_{\alpha}} (\bK , t) ,
\end{equation}
where ${ C_\ell (\bK, t) }$ is a function decaying like a Gaussian
\begin{equation}
  \label{eq:C_bath_exp}
  C_\ell (\bK, t) = n (\bK) \, \re^{-\frac{A_{\ell}}{2}{(t/\Tc(\bK))}^2},
\end{equation}
where we introduced the torque time
\begin{equation}
\Tc (\bK) = \frac{1}{\Gamma (\bK)} .
\label{def_TC}
\end{equation}

Equations~\eqref{eq:C_bath} and~\eqref{eq:C_bath_exp} are the main result of
this section, as they provide us with a simple estimate for the time evolution
of the ensemble-averaged correlation function of the fluctuations in the system. In
Fig.~\ref{fig:CorrelationOphi} we compare this estimate to the correlations
measured in the $N$-body simulations, and to shorten the main text,
we detail the procedure followed to obtain that figure in Appendix~\ref{sec:WindowOphi}.
\begin{figure}
\begin{center}
\includegraphics[width=0.48\textwidth]{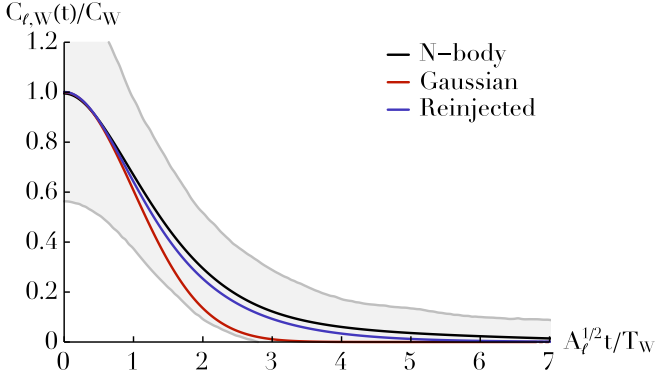}
\caption{ Correlation of the noise fluctuations, ${C_{\ell , W} (\bK, t)}$, for
${ \ell \!=\! 2 }$ averaged over a window in $\bK$ such that ${ (C_{\mathrm{min}} , \Tc^{\mathrm{min}}) \!\simeq\! (5.0 , 142) }$ with ${ \veps_{W} \!=\! 0.1 }$, as defined in Eq.~\eqref{def_W},
for which there are on average $9$ particles in the window per simulation.
The
typical amplitude and torque time are given by ${ (C_{W} , T_{W}) \!\simeq\! (0.72 , 148) }$ (see Eq.~\eqref{def_oCW_TW}).
The
black line was ensemble-averaged over ${1000}$ realizations of the fiducial
system. The background gray lines illustrate the 10\% and 90\% spreads over
these realizations. The red line is the Gaussian prediction from
Eq.~\eqref{Approximation_Correlation_Bath}.
The purple line is the updated prediction obtained by reinjecting the Gaussian prediction
into the self-consistency relation from Eq.~\eqref{self_relation}, as detailed in Eq.~\eqref{reinjected_prediction},
that decays exponentially at late times.
\label{fig:CorrelationOphi}}
\end{center}
\end{figure}
As expected, this estimation matches the $N$-body
measurements on short timescales.  Capturing the late-time non-Gaussian
behavior of the correlation function requires a self-consistent determination
of the time-dependence of the noise.
This is investigated in Section~\ref{sec:SelfConsistency}, and allows for
an improved noise prediction in Fig.~\ref{fig:CorrelationOphi}.
Here, such a calculation is made intricate by our accounting of
the $\ell$- and $\bK$-dependence of
the pairwise coupling, as embodied by the sum over $\ell$ and the integral over ${ \rd \bKp }$
in Eq.~\eqref{EOM_vphi_II}.

In Section~\ref{sec:TestParticle}, we will use the previous
correlation functions as source terms to describe the dynamics of a test
particle embedded in that noisy environment.
However, in that section we will see that the ensemble averaged correlation function
does not capture the full dynamics induced on a test particle.
Indeed, globally conserved quantities (such as the total energy) prevent the system from being fully ergodic:
even after long times the system will not explore the entire realization space
and therefore time averages are not equivalent to ensemble averages.
For a given realization `$\rr$', we therefore define the time-averaged correlation function
\begin{equation}
\label{def_C_real}
C_{\alpha \beta}^{\rr} (\bK , \bK' , t - \tp) \equiv \avT{ \vphi_{\alpha} (\bK , t) \, \vphi_{\beta} (\bK' , \tp) } ,
\end{equation}
where
\begin{equation}
\savT{ f } \equiv \frac{1}{T} \!\! \int_{0}^{T} \!\!\!\! \rd t \, f (t)
\label{def_time_average}
\end{equation}
stands for the time average over some long timescale $T$.
As previously, we will assume that $C_{\alpha \beta}^{\rr} (\bK , \bK' , t)$
can be replaced by
\begin{equation}
\label{Gaussian_Cbath}
C_{\alpha\beta}^{\rr} (\bK , \bK' , t) = \delta_{\alpha}^{\beta} \, \deltaD (\bK - \bK') \, C_{\ell_\alpha}^{\rr} (\bK , t) ,
\end{equation}
with the Gaussian time dependence
\begin{equation}
\label{def_oC}
C_{\ell}^{\rr} (\bK , t) = n_{\ell}^{\rr} (\bK) \, \re^{-\frac{A_{\ell}}{2}{(t/\Tc(\bK))}^2} .
\end{equation}
Here, we defined an effective isotropic amplitude ${ n_{\ell}^{\rr} (\bK) }$ as the mean value over $m_{\alpha}$, so that
\begin{align}
\label{def_oC_m}
n_{\ell_{\alpha}}^{\rr} (\bK) \!=\! \frac{1}{2 \ell_{\alpha} + 1} \sum_{m_{\alpha} , \beta} \! \int \! \rd \bK_{\beta} \, C_{\alpha \beta}^{\rr} (\bK_{\alpha} , \bK_{\beta} , 0) ,
\end{align}
and from Eqs.~\eqref{C0_bath}, we have
\begin{align}
\label{def_C0}
\big\langle n_{\ell}^{\rr} (\bK) \big\rangle  & = n (\bK),
\end{align}
that is independent of the considered harmonic.  It is important to note that
${ n_{\ell}^{\rr} (\bK) }$ varies between different
realizations.  In Eq.~\eqref{var_oC}, we illustrate how one can compute its
variance, and show how this originates from the constraint of total energy
conservation.

In Eq.~\eqref{def_oC}, we assumed, for simplicity, that the torque time,
${ \Tc(\bK) }$, is independent of the considered realization.
These various choices ensure that the ansatz from Eq.~\eqref{Gaussian_Cbath}
satisfies the constraints from Eqs.~\eqref{C0_bath}
and~\eqref{calc_C2_III_bath} when ensemble-averaged.  As highlighted by
Eqs.~\eqref{Gaussian_Cbath}, this correlation is diagonal both w.r.t.\ the
harmonic indices (via $\delta_{\alpha}^{\beta}$) and w.r.t.\ the considered
parameters (via ${ \deltaD (\bK - \bK') }$).  Following
Eq.~\eqref{def_oC}, we note that the time-dependence of this correlation is
controlled by both an isotropic amplitude,
${ n_{\ell}^{\rr} (\bK) }$, and a torque time,
${ \Tc (\bK) }$, that depend on the considered parameter $\bK$.

\section{The random walk of a test particle}
\label{sec:TestParticle}

\begin{figure}
\begin{center}
\includegraphics[width=0.35\textwidth]{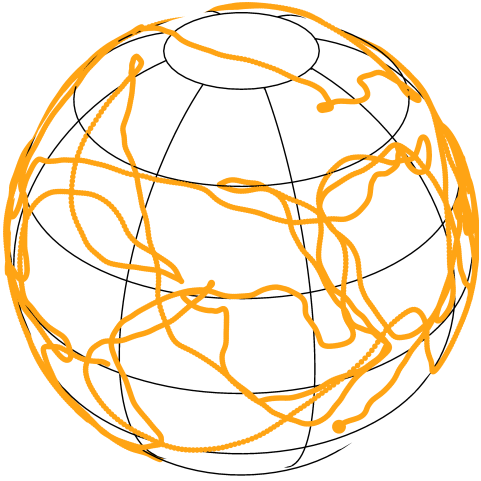}
\caption{
Illustration of the random walk in orientation of a given test particle from the fiducial simulations, following the same convention as in Fig.~\ref{fig:IllustrationSystem}, and represented for ${ 0 \!\leq\! t \!\leq\! 10^{4} \!\times\! 20 h }$. 
In Section~\ref{sec:TestParticle}, we characterize the statistical properties of that random walk on the sphere.
\label{fig:IllustrationRandom}}
\end{center}
\end{figure}

In the previous section, we characterized the noise fluctuations resulting
from the coupled motions of the system's $N$ particles.
Assuming that the
statistics of this noise follows the correlation function obtained in
Eq.~\eqref{eq:C_bath}, our goal is now to investigate the
stochastic dynamics of one given test particle embedded in that fluctuating
environment. In Fig.~\ref{fig:IllustrationRandom}, we illustrate one such
random walk by highlighting the time evolution of the orientation of a single
particle in one fiducial simulation.

Throughout this section, we use the test particle limit, i.e., we assume that the
motion of the test particle is fully determined by the time-dependent density of
the background particles and we neglect any backreaction of the test particle onto
the background particles.  We denote the parameters of the test particle with
$\bKt$, and its orientation at time $t$ with ${ \bLt (t) }$.  Similarly to
Eq.~\eqref{def_vphi}, the current orientation of the test particle is fully
characterized by the single particle \DF\
\begin{equation}
\vphit (\bL , t) = \deltaD (\bL - \bLt (t)) ,
\label{def_vphit}
\end{equation}
which can be expanded as
${ \vphit (\bL , t) = Y_{\alpha} (\bL) \, \vphit_{\alpha} (t) }$ (with the sum
over $\alpha$ implied), where
\begin{equation}
\vphit_{\alpha} (t) = Y_{\alpha} (\bLt (t)).
\label{def_vphit_alpha}
\end{equation}
Since
${ \vphit_{(1 , m)} (\bLt) \propto ( y_{\rt} , z_{\rt} , x_{\rt} ) \propto \bLt
}$, characterizing the random walk of the test particle on the sphere as in
Fig.~\ref{fig:IllustrationRandom}, requires the knowledge of the correlation
properties of ${ \vphit_{\alpha} (t) }$ for ${ \ell_{\alpha} = 1 }$.

The evolution equation for ${ \vphit_{\alpha} (t) }$ follows from
Eq.~\eqref{EOM_vphi_II}, and reads
\begin{align}
  \frac{\partial \vphit_{\alpha} (t)}{\partial t} 
  &
    \, = - \!\! \int \!\! \rd \bK \, Q_{\alpha \gamma \delta} (\bKt , \bK) \, \vphi_{\gamma} (\bK , t) \, \vphit_{\delta} (t)
    \nonumber
  \\
  & \, = - \Qt_{\alpha \delta} (t) \, \vphit_{\delta} (t) ,
\label{evol_vphit}
\end{align}
where ${ \vphi_{\gamma} (\bK , t) }$ is the harmonic coefficients of the
background particles' field, as defined in Eq.~\eqref{def_vphi_alpha}, and we
introduced
${ \Qt_{\alpha \delta} (t) \!=\! \!\int\! \rd \bK \, Q_{\alpha \gamma \delta} (\bKt , \bK) \,
  \vphi_{\gamma} (\bK , t) }$, that is a time-dependent external forcing term
driving the dynamics of the test particle. The time-dependence of this
source only originates from the background particles (via
${ \vphi_{\gamma} (\bK , t) }$), whose correlation properties were investigated
in the previous section.
We also emphasize that this forcing term also depends on $\bKt$, the parameters
of the considered test particle.

Equation~\eqref{evol_vphit} takes the form of a time-dependent linear matrix
differential equation for the test particle's harmonic coefficients
$\vphit_{\alpha} (t)$.  In order to guarantee well-behaved asymptotics of the
test particle's motion for large times, we approach the resolution of
Eq.~\eqref{evol_vphit} via Magnus series (see~\cite{Blanes2009} for a review).
In that framework, one can generically solve for the motion of the test
particle as
\begin{equation}
\label{Magnus_solution}
  \vphit_{\alpha} (t) = {\big[ \re^{\Omega (\tp , t)} \big]}_{\alpha \delta} \, \vphit_{\delta} (\tp) ,
\end{equation}
where the matrix ${ \Omega (\tp , t) }$ is constructed as a series expansion of the form ${ \Omega (\tp , t) = \sum_{k \geq 1} \Omega_{k} (\tp , t) }$, whose first terms are
\begin{align}
\Omega_{1} (\tp , t) & \, = - \!\! \int_{\tp}^{t} \!\! \rd s \, \Qt (s) ,
\nonumber
\\
\Omega_{2} (\tp , t) & \, = \frac{1}{2} \!\! \int_{\tp}^{t} \!\! \rd s_{1} \!\! \int_{\tp}^{s_{1}} \!\! \rd s_{2} \, \big[ \Qt (s_{1}) , \Qt (s_{2}) \big] ,
\label{series_O}
\end{align}
where ${ [ A , B ] = A B \!-\! B A }$ is the matrix commutator.
As in Eq.~\eqref{def_C_real}, for a given realization, the statistics of the motion
of the test particle are captured by the stationary time-averaged correlation function
\begin{equation}
C^{\rt , \rr}_{\alpha \beta} (t - \tp) \equiv \avT{ \vphit_{\alpha} (t) \, \vphit_{\beta} (\tp) } .
\label{def_Ct}
\end{equation}
Using Eq.~\eqref{Magnus_solution}, we can write this correlation function as
\begin{align}
C^{\rt , \rr}_{\alpha \beta} (t) = \avT{{\big[ \re^{\Omega (t)} \big]}_{\alpha \delta}} \, \avLt{\vphit_{\delta} (0) \, \vphit_{\beta} (0)} ,
\label{calc_Ct}
\end{align}
where we relied on our test particle's assumption (i.e.\ independence hypothesis~\citep{Corrsin1959}),
which allowed us to separate the time average (denoted with ${ \savT{ \, \cdot \, } }$) over the background particles generating the noise, and the average over the initial location of the test particle (denoted with ${ \savLt{ \, \cdot \, } }$).
We also relied on the hypothesis that the noise is stationary in time, so that ${ \savT{\re^{\Omega (\tp , t)}} = \savT{\re^{\Omega (t - \tp)}} }$, with ${ \Omega (t) \equiv \Omega (0, t) }$.

In Appendix~\ref{sec:CalcRandomWalk}, we rely on the cumulant theorem to compute the two averages appearing in Eq.~\eqref{calc_Ct}.
It allows us to rewrite the test's particle correlation function as
\begin{align}
  C^{\rt , \rr}_{\alpha \beta} (t) = 
  & \, \frac{\delta_{\alpha}^{\beta}}{4 \pi} \, \exp \!\bigg\{ -
    A_{\ell_{\alpha}} \sum_{\ell} B_{\ell}
    \!\! \int \!\! \rd \bK \, n_{\ell}^{\rr} (\bK)
    \mJ_{\ell}^2 \big[ \bKt, \bK \big]
    \nonumber  \\
  &\;\;\;\;\;\;\;\;\;\;\;\;\;\;\;\;  \, \times  
    \frac{\Tc^2(\bK)}{A_{\ell}}\chi\bigg[
    \frac{t\sqrt{A_{\ell}/2}}{\Tc(\bK)}\bigg]
    \bigg\}
  \label{final_Ct}
 \\
    & \hspace{-1.5cm} \simeq \frac{\delta_{\alpha}^{\beta}}{4 \pi} \!\times\!\!
    \begin{cases}
    \!\re^{- \frac{A_{\ell_{\alpha}}}{2} \big(\! \sum_{\ell}\! B_{\ell} \!\!\int\!\! \rd \bK \, n_{\ell}^{\rr} (\bK) \mJ_{\ell}^{2} [ \bK_{\rt} , \bK ] \!\big) t^{2} } &  \hspace{-0.3cm} t \ll \Tc ,
    \\
     \!\re^{- \frac{\sqrt{\smash[b]{\pi}} A_{\ell_{\alpha}}}{2} \big(\! \sum_{\ell}\! \frac{B_{\ell}}{\sqrt{\smash[b]{A_{\ell} \!/ 2}}} \!\!\int\!\! \rd \bK \, n_{\ell}^{\rr} (\bK) \mJ_{\ell}^{2} [ \bK_{\rt} , \bK ] \Tc (\bK) \!\big) t } & \hspace{-0.3cm} t \gg \Tc ,
    \end{cases}
    \nonumber
\end{align}
where we introduced the dimensionless function
\begin{align}
\label{eq:def_chi}
\chi(\tau) & =  \!\! \int_{0}^{\tau} \!\! \rd \tau_{1} \!\! \int_{0}^{\tau} \!\! \rd \tau_{2} \, \re^{ - (\tau_{1} - \tau_{2})^{2}}
\nonumber
\\
& = - 1 + \re^{- \tau^{2}} \!+\! \tau \, \sqrt{\pi} \, \erf ( \, \tau)
\nonumber
\\
& \, \simeq
\begin{cases}
\tau^2 & \tau \ll 1 ,
\\
\sqrt{\pi} \tau & \tau \gg 1.
\end{cases}
\end{align}

As can be seen from the time-dependence of the exponent in \eq~\eqref{final_Ct},
one can note that on short timescales,
${ t \ll \Tc }$, the correlation $C^{\rt , \rr}_{\alpha \beta}$ decays like a
Gaussian and the motion of the particle is ballistic (e.g.,
${ (\Delta \bL)^{2} \propto t^{2} }$). On that short timescales, the random walk of the test
particle is analogous to the one induced by a time-independent fluctuation.
On long timescales,
${ t \gg \Tc }$, the correlation decays exponentially in time and the motion of
the test star is diffusive (e.g.,
${ (\Delta \bL)^{2} \propto t }$). On these long timescales, the random walk of the
test particle is analogous to the one induced by fluctuations
$\deltaD$-correlated in time (as in the classical Brownian motion), leading to a
diffusive random walk on the sphere.

Because it involves an integral over $\bK$, Eq.~\eqref{final_Ct} remains
difficult to implement.
Let us now present a simpler toy model to generate a stochastic motion on the sphere
that would share correlation properties similar to the ones of Eq.~\eqref{final_Ct}.
As such, we will assume that the stochastic motion of the test particle is generated by
an effective dipole Gaussian noise, and follows the Langevin equation
\begin{equation}
  \label{eq:dipole_noise}
  \frac{\rd \bL}{\rd t} = \Gamma_{\rt} \, \bmeta (t) \times \bL,
\end{equation}
where the Gaussian noise ${ \bmeta (t) }$ is a ${3D}$ vector of zero mean,
${ \savEA{\eta_{i} (t)} = 0 }$,
and follows ${ \savEA{\eta_i(t) \, \eta_j(t')} = \delta_{ij} \, \re^{-{[(t-t')/\Tc^\rt]}^2} }$.
We will then choose the amplitude $\Gamma_{\rt}$ and
coherence time $\Tc^{\rt}$ by matching the short and long timescales
behavior of the test particle's correlation function with the ballistic and
diffusive regimes of the generic result from Eq.~\eqref{final_Ct},
an approach already used in~\cite{Hamers2018}.

Following the same steps as in Eq.~\eqref{Magnus_solution},
we may compute the correlation function of a test particle,
whose dynamics is imposed by Eq.~\eqref{eq:dipole_noise}.
It reads
\begin{align}
  \label{Ct_dipole}
  C^{\rt , \rr}_{\alpha \beta} (t) & \, = 
  \frac{\delta_{\alpha}^{\beta}}{4 \pi} \,
    \re^{- \frac{ A_{\ell_{\alpha}}}{2} \Gamma_\rt^2 (\Tc^\rt)^2\chi(t/\Tc^\rt)}
    \nonumber \\ 
  & \simeq \frac{\delta_{\alpha}^{\beta}}{4 \pi} \times  \begin{cases}
\,  \, \re^{ - \frac{A_{\ell_{\alpha}}}{2} \Gamma_\rt^2t^2} & t \ll \Tc^\rt ,
\\
\re^{ -  \frac{ \sqrt{\pi} A_{\ell_{\alpha}}}{2}  \Gamma_\rt^2 \Tc^\rt t} & t \gg \Tc^\rt ,
\end{cases}
\end{align}
with ${ A_{\ell} = \ell (\ell + 1) }$.
By matching the ballistic and diffusive regimes of Eq.~\eqref{Ct_dipole}
with the ones of Eq.~\eqref{final_Ct}, we may then constrain
the amplitude, $\Gamma_\rt$, and coherence time, $\Tc^\rt$,
of the toy model of Eq.~\eqref{eq:dipole_noise}.

The amplitude, ${ \Gamma_{\rt} (\bK_{\rt}) }$, varies from realization to realization and is given by
\begin{align}
\label{def_kappa}
\Gamma_\rt^2(\bKt) & = \sum_{\ell} B_{\ell} \!\! \int \!\! \rd \bK \,
                     n_{\ell}^{\rr} (\bK) \, \mJ_{\ell}^2 \big[ \bKt , \bK \big],
\end{align}
When ensemble-averaged over realizations, this amplitude becomes
\begin{align}
\label{kappa_mean}
\langle \Gamma_\rt^2(\bKt) \rangle = \Gamma^2 (\bKt),
\end{align}
as already defined in Eq.~\eqref{Gamma_bath}.
Finally, the coherence time is given by
\begin{equation}
  \label{eq:Tct_def}
\langle \Gamma_\rt^2 (\bK_{\rt}) \rangle \, \Tc^{\rt}(\bKt)  \!=\!  \!\sum_{\ell}\!
            \frac{B_{\ell}}{\sqrt{\smash[b]{A_{\ell}/2}}} \! \int \!\! \rd
            \bK \,  
            n (\bK)
            \mJ_{\ell}^2 \big[ \bKt , \bK \big] \, \Tc (\bK) ,
\end{equation}
where for simplicity we assumed, similarly to Eq.~\eqref{def_oC},
that the coherence time, ${ \Tc^{\rt} (\bK_{\rt}) }$, is independent of the considered realization.

\begin{figure}
\begin{center}
\includegraphics[width=0.35\textwidth]{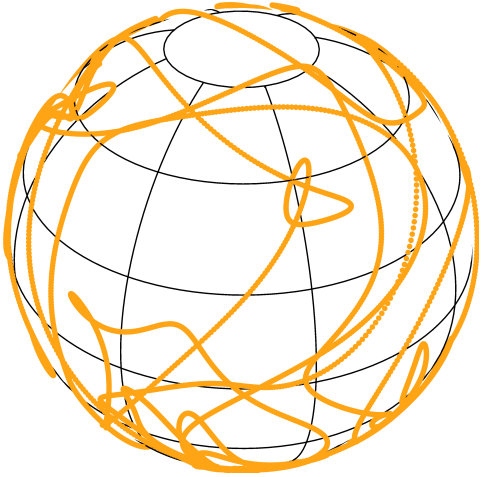}
\caption{
Illustration of a random walk generated by the stochastic equation~\eqref{eq:dipole_noise} following the same convention as in Fig.~\ref{fig:IllustrationRandom} and considering a test particle with the same $\bK_{\rt}$ parameters.
\label{fig:IllustrationRandomFake}}
\end{center}
\end{figure}

Equation~\eqref{Ct_dipole} is the key result of this section.  Indeed, it
provides us with an analytical description of the statistical properties of the
random walk of a test particle's orientation, as jointly induced by the
fluctuating noise from the background particles.  The test particle's random
walk is characterized by the two quantities
${ (\Gamma_\rt^{2} , \Tc^{\rt}) }$, that both depend on the test
particle's parameters $\bKt$. The coefficient ${ \Gamma_\rt }$
controls the amplitude of the test particle's initial ballistic motion, while
the coherence time ${ \Tc^{\rt} }$ controls the timescale after which
the test particle enters the diffusive
regime.
One strength of the
present formalism is that, following Eqs.~\eqref{def_kappa} and~\eqref{eq:Tct_def},
one now has at one's disposal explicit expressions for these two parameters.
These coefficients can then easily be computed for various cluster models (by varying the \DF\
${ n(\bK) }$) and various test particles (by varying $\bKt$).
Considering the same test particle as in Fig.~\ref{fig:IllustrationRandom},
we illustrate in Fig.~\ref{fig:IllustrationRandomFake} one random walk generated
using the Langevin equation~\eqref{eq:dipole_noise}.

However, as we had already emphasized in Eq.~\eqref{Gaussian_Cbath},
it is important to note that the present system suffers from being
non-ergodic, i.e.\ ensemble averages and time averages cannot be interverted.
This is highlighted by the fact that even for the exact same test particle
(i.e.\ same $\bK_{\rt}$), the amplitude ${ \Gamma_\rt }$
varies from realization to realization.
In Appendix~\ref{sec:NoiseVariance}, we compute the associated variance,
as given by Eq.~\eqref{var_Ct}, and show that this effect originates from the constraint
of total energy conservation.
Moreover, we show that this variance remains non-zero even in the limit of a
Gaussian noise, and as such does not vanish in the limit of an infinite number
of background particles generating the fluctuations.

Let us finally use our fiducial numerical simulations to highlight the result
from Eq.~\eqref{Ct_dipole}. This is illustrated in Fig.~\ref{fig:RandomWalk},
and to shorten the main text, we detailed in
Appendix~\ref{sec:WindowRandomWalk} the procedure followed to obtain that
figure.
\begin{figure}
\begin{center}
\includegraphics[width=0.48\textwidth]{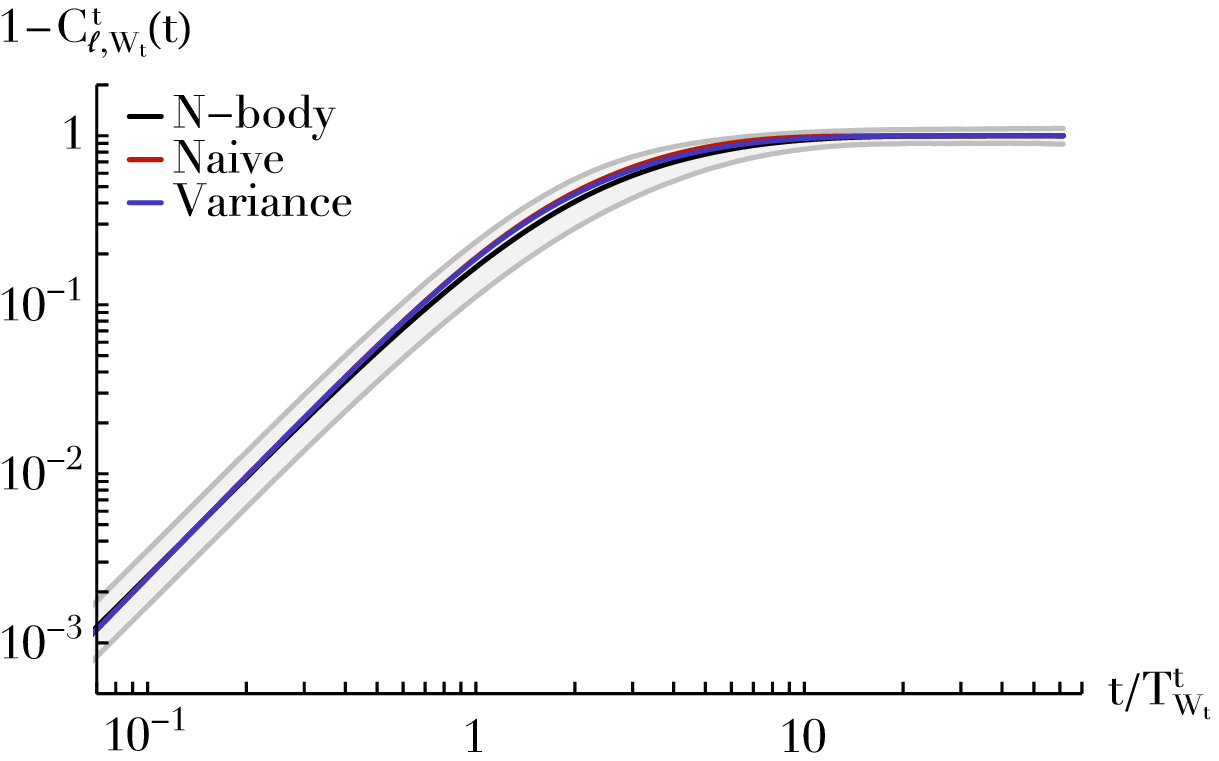}
\includegraphics[width=0.48\textwidth]{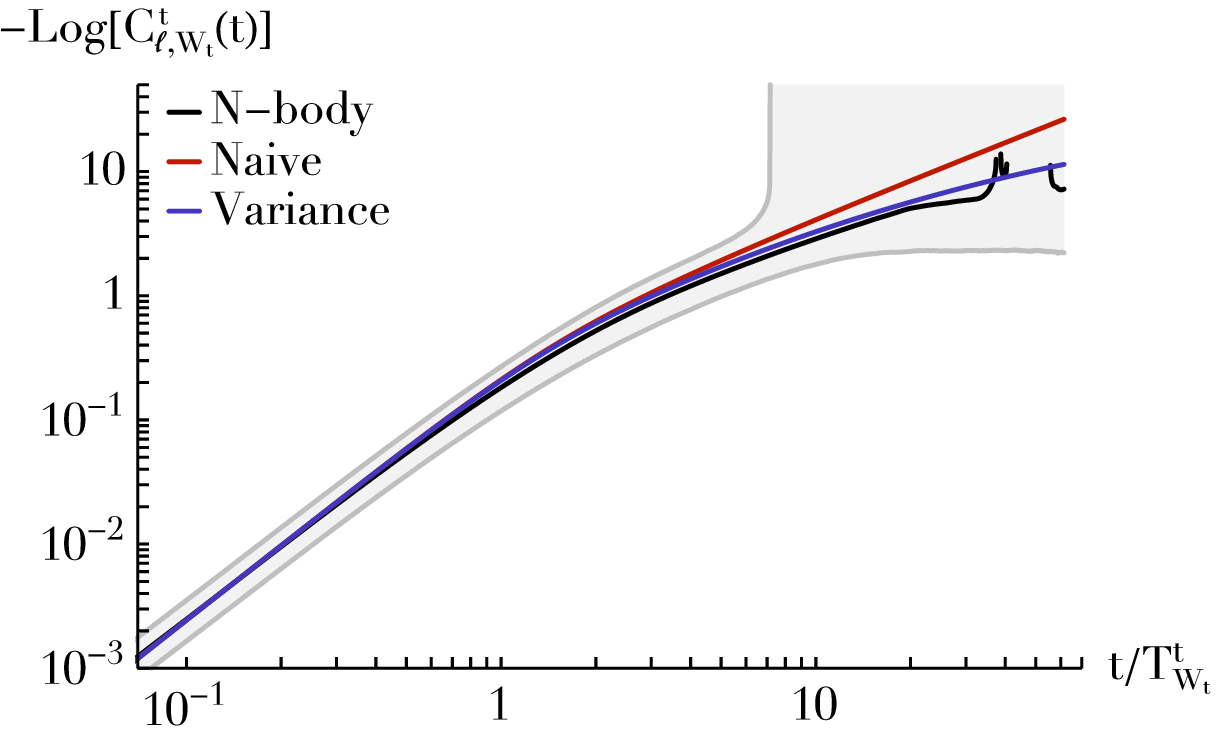}
\caption{Illustration of the correlated random walk of test particles, as captured by the correlation function ${ C_{\ell , W_{\rt}}^{\rt} (t) }$ for ${ \ell =1 }$, following the definition from Eq.~\eqref{def_Ct_mean}.
The window function ${ W_{\rt} (\bKt) }$ (see Eq.~\eqref{def_Wt}) is characterized by ${ ( \Gamma_{\rm min}^{2}, T_{\rt}^{\rm min}) = (1.2 \!\times\! 10^{-4} , 40.0) }$ and ${ \veps_{W_{\rt}} = 0.1 }$.
The black line was ensemble-averaged over ${1000}$ realizations of the fiducial system.
The background gray lines illustrate the 10\% and 90\% spreads over these realizations.
The red line follows the naive approximation from Eq.~\eqref{Naive_Ct_mean}, for which one finds ${ (\Gamma_{W_{\rt}}^{2} , T_{W_{\rt}}^{\rt}) \simeq (1.3 \!\times\! 10^{-4} , 43.6) }$ (see Eq.~\eqref{Naive_Ct_other}).
The blue line follows the approximation from Eq.~\eqref{variance_Ct_mean}, which accounts for the variance of ${ \Gamma_{\rt}^{2} }$.
In the top panel, one can note the initial ballistic regime and the subsequent saturation of the diffusion.
The bottom panel also illustrates the two successive regimes, namely ballistic (${ \propto t^{2} }$ for ${ t \ll T_{W_{\rt}}^{\rt} }$) followed by diffusive (${ \propto t }$ for ${ t \gg T_{W_{\rt}}^{\rt} }$), as emphasized in Eq.~\eqref{final_Ct}.
This panel also highlights the importance of accounting for the variance in the amplitude $\Gamma_{\rt}^{2}$
to correct the late-time behavior of the test particles' random walks.
\label{fig:RandomWalk}}
\end{center}
\end{figure}
In that figure, we note that the $N$-body measurements and the analytical
prediction from Eq.~\eqref{Ct_dipole} agree both on short timescales but also
on timescales longer than the coherence time $T_{W_{\rt}}^{\rt}$ (defined in
Eq.~\eqref{Naive_Ct_other}).  As already stressed in Eq.~\eqref{final_Ct}, the
second panel of Fig.~\ref{fig:RandomWalk} clearly exhibits the two successive
regimes of evolution, namely ballistic for ${ t \ll \Tc^{\rt} }$ and
diffusive for ${ t \gg \Tc^{\rt} }$.  This same panel also emphasizes the
importance of accounting for the variance in ${ \Gamma_{\rt}^{2} }$, to correctly capture the
late-time behavior of the test particles' stochastic motions.  We recall that
this effect that does not vanish in the limit of an infinite number of
background particles. Since ${ \vphit_{(1,m)} (t) \propto \bL_{\rt} }$,
Fig.~\ref{fig:RandomWalk} also offers then an illustration of the behavior of
${ t \mapsto \savEA{\bLt (t) \, \bLt (0)} }$.
It is also straightforward to
adapt that prediction to different test stars (by changing $\bKt$) or different
galactic nuclei (by changing the \DF\ ${ n (\bK) }$).

\section{The self-consistency of the noise}
\label{sec:SelfConsistency}

In the previous derivations, we proceeded in two successive steps.
First, in Section~\ref{sec:CorrelNoise}, we used estimates of the derivatives of
the correlation function of the noise
at the initial time to obtain an ansatz in Eq.~\eqref{eq:C_bath_exp} for the time-dependence
of the correlation function of the noise generated by the $N$ background particles.
Then, in Section~\ref{sec:TestParticle}, we used this noise as a source term to study
the stochastic dynamics of a test particle.
Yet, if the considered test particle is taken to be one particular background particle,
its random walk in orientation and the background fluctuations sourcing it have to satisfy
some self-consistency relation. This is what we explore in this section.

We start from Eq.~\eqref{def_C_real}, and replace ${ \vphi_{\alpha} (\bK , t) }$
by its definition in terms of a discrete sum over particles, as in Eq.~\eqref{def_vphi_alpha},
so that
\begin{align}
C_{\alpha\beta}^{\rr} (\bK , \bKp , t - \tp) = & \, \sum_{i , j} \deltaD (\bK - \bK_{i}) \, \deltaD (\bKp - \bK_{j})
\nonumber
\\
&\, \times \avT{Y_{\alpha} (\bL_{i} (t)) \, Y_{\beta} (\bL_{j} (\tp))} .
\label{calc_sc}
\end{align}
We now assume that each background particle can be treated as a test particle,
and that their long-term motions are decorrelated one from another.
Only contributions from ${ i = j }$ remain, and Eq.~\eqref{calc_sc} becomes
\begin{equation}
C_{\alpha \beta}^{\rr} (\bK , \bKp , t ) = \deltaD (\bK - \bKp) \sum_{i} \deltaD (\bK - \bK_{i}) \, C_{\alpha\beta}^{\rt , \rr} (\bK_{i} , t) ,
\label{calc_sc_II}
\end{equation}
where ${ C_{\alpha\beta}^{\rt , \rr} (\bK_{i} , t) }$ is the test particle's correlation function of the particle $i$, as defined in Eq.~\eqref{def_Ct}.
To proceed further, let us now take the ensemble average of both sides of Eq.~\eqref{calc_sc_II}, to get
\begin{equation}
C_{\alpha \beta} (\bK , \bKp , t) = \deltaD (\bK - \bKp) \, 4 \pi \, n (\bK) \, \avEA{C_{\alpha \beta}^{\rt , \rr} (\bK , t)} .
\label{calc_sc_III}
\end{equation}

Luckily, in Eq.~\eqref{calc_Ct}, we have already solved for the correlation function
of the random walks of the test particle through Magnus series.
Using Eq.~\eqref{calc_cumulant_II}, we generically get
\begin{align}
C_{\alpha \beta}^{\rt , \rr} (\bK , t) = \frac{\delta_{\alpha}^{\beta}}{4 \pi} \, & \, \exp \bigg\{ - \frac{A_{\ell_{\alpha}}}{2} \,  \sum_{\ell} B_{\ell} \!\! \int \!\! \rd \bKp \, \mJ_{\ell}^{2} \big[ \bK , \bKp \big]
\nonumber
\\
& \, \times \!\! \int_{0}^{t} \!\! \rd t_{1} \!\! \int_{0}^{t} \!\! \rd t_{2} \, C_{\ell}^{\rr} (\bKp , t_{1} - t_{2}) \bigg\} .
\label{solution_sc}
\end{align}
We may then take the ensemble-average of this relation, and for simplicity,
keep only the first cumulant in the cumulant theorem.
Reinjected into Eq.~\eqref{calc_sc_III}, this leads to
\begin{align}
C_{\ell} (\bK , t) = n (\bK) & \, \exp \bigg\{ - \frac{A_{\ell}}{2} \sum_{\ellp} B_{\ellp} \!\! \int \!\! \rd \bKp \, \mJ_{\ellp}^{2} \big[ \bK , \bKp \big]
\nonumber
\\
& \, \times \!\! \int_{0}^{t} \!\! \rd t_{1} \!\! \int_{0}^{t} \!\! \rd t_{2} \, C_{\ellp} (\bKp , t_{1} - t_{2})  \bigg\} .
\label{self_relation}
\end{align}
Equation~\eqref{self_relation} takes the form a self-consistent integral equation satisfied by the correlation of the noise fluctuations in the system.
This relation can be further clarified by defining
\begin{equation}
R_{\ell} (\bK , t) = \frac{1}{2} \! \int_{0}^{t} \!\! \rd t_{1} \!\! \int_{0}^{t} \!\! \rd t_{2} \, C_{\ell} (\bK , t_{1} - t_{2}) ,
\label{def_Rell_sc}
\end{equation}
and one finally gets the self-consistent differential equation\footnote{Similar self-consistent differential equations were first obtained in~\cite{TaylorMcNamara1971} (see Eq.~{(22)} therein) in the context of plasma diffusion in the guiding center limit.}
\begin{align}
\frac{\rd^{2} R_{\ell} (\bK , t)}{\rd t^{2}} = n (\bK) \, \exp \bigg\{ & - A_{\ell} \sum_{\ellp} B_{\ellp} \!\! \int \!\! \rd \bKp \, \mJ_{\ellp}^{2} \big[ \bK , \bKp \big]
\nonumber
\\
& \, \times \, R_{\ellp} (\bKp , t) \bigg\} .
\label{selfequation_Rell}
\end{align}
Equation~\eqref{selfequation_Rell} is the important result of this section,
as it highlights the self-consistent relation satisfied by the correlation of the noise fluctuations.
Yet, as it couples both different harmonics (via $\sum_{\ellp}$) and different parameters
(via ${ \!\int\! \rd \bKp }$), such a differential equation appears too intricate to
easily be solved explicitly. This is not pursued further here.

One may still proceed iteratively to obtain improved approximations
of the noise correlation function.
To do so, one starts from the Gaussian dependence obtained in Eq.~\eqref{eq:C_bath_exp}.
This (motivated) ansatz may then be reinjected in the r.h.s.\ of the 
self-consistency relation from Eq.~\eqref{self_relation}, leading to a new expression
of the noise correlation function that would have both a ballistic and a diffusive part.
Such a procedure is illustrated in Fig.~\ref{fig:CorrelationOphi}, where we show
how one can better match the late-time properties of the system's noise
through this iterative process.

\section{Application}
\label{sec:PowerLaw}

As an illustration of the present formalism, let us now consider the case of a stellar cusp
distribution similar to the one of SgrA*.
The mass of the \MBH\ is taken to be ${ \mBH \!=\! 4 \!\times\! 10^{6} M_{\odot} }$,
and for simplicity we consider a single-mass stellar population of individual mass ${ m_{\star} \!=\! 1 M_{\odot} }$.
We assume that the stars' eccentricities follow a thermal distribution,
${ f_{e} (e) \!=\! 2 e }$~\citep{Merritt2013},
and that the number of stars per unit $a$ follows a power law distribution of the form
${ n_{a} (a) \!=\! (N_{0} / a_{0}) {(a / a_{0})}^{2 - \gamma} }$,
where ${ N_{0} \!=\!  g(\gamma) N( \!<\! a_{0}) }$,
with 
\begin{equation}
  \label{eq:g_gamma}
  g (\gamma) \!=\! 2^{- \gamma} \, (3 - \gamma) \, \sqrt{\smash[b]{\pi}} \, \frac{\Gamma (1 + \gamma)}{\Gamma (\gamma - \tfrac{1}{2})}
\end{equation}
and ${ N ( \!<\! a_{0}) }$ the physical number of stars within a sphere of
radius $a_{0}$ from the center.
For the numerical application, we assume that ${ a_{0} \!=\! \rh \!=\! 2 \,\pc }$ and ${ N (\!<\! a_{0}) \!=\! 4 \!\times\! 10^{6} }$.
The system being of infinite extent, we write the system's \DF\ as
${ n (m , a , e) \!=\! f_{m} (m) f_{e} (e) \, n_{a} (a) / (4 \pi) }$.

\begin{figure}
\begin{center}
\includegraphics[width=0.47\textwidth]{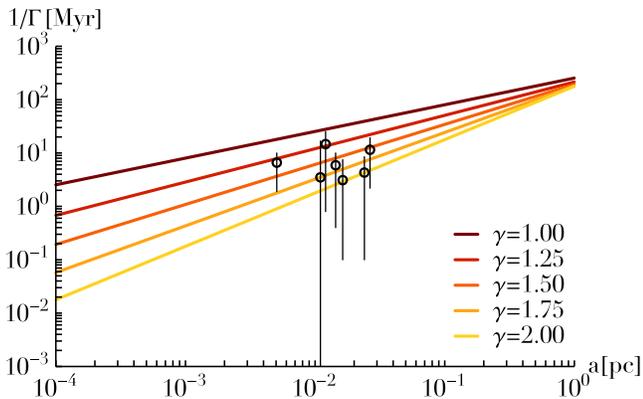}
\caption{
Illustration of the torque time, ${ 1/\Gamma }$, for circular orbits (${ e = 0 }$) as a function of the semi-major axis, and for different cusp's profiles (through the power index $\gamma$) similar to the one around SgrA*.
For comparison, black circles with errors show the main sequence ages of some
of the S-stars~\citep{Habibi2017}.
\label{fig:PowerLawInvGamma}}
\end{center}
\end{figure}

In Appendix~\ref{sec:CalcPowerLaw}, we show that the amplitude ${ \Gamma^{2} }$,
defined in Eq.~\eqref{Gamma_bath}
and characterizing the ballistic regime of the orientation's random walk,
follows the power law distribution
\begin{equation}
\Gamma^{2} (a , e) = \frac{N( \!<\! a)}{P^2(a)} \frac{\langle m^2 \rangle}{M_\bullet^2}
\, \frac{\pi g(\gamma) f_{\Gamma^{2}} (e)}{1 - e^{2}} \, ,
\label{res_Gamma_pwl}
\end{equation}
where ${ P(a) = 2 \pi (a^{3} / (G \mBH))^{1/2} }$ is the orbital period, and ${ f_{\Gamma^{2}} (e) \simeq 0.15 }$
is a dimensionless eccentricity function defined in
Eq.~\eqref{def_Gamma0_fGamma_pwl} and illustrated in Fig.~\ref{fig:PowerLawf}.
In Fig.~\ref{fig:PowerLawInvGamma}, we illustrate the dependence of the torque
time, ${ 1/ \Gamma }$, for circular orbits of different semi-major axes and
for different cusp's profiles, and interestingly note that this \VRR\ timescale
is similar to the age of some of the young stars observed in our Galactic
center~\citep{Habibi2017}.
As shown in Fig.~\ref{fig:PowerLawInvGamma}, should the S-stars be born in
a disk, the \VRR\ process is sufficiently fast to isotropize their orbital orientations~\citep{HopmanAlexander2006},
but \SRR\ may not be efficient enough to thermalize
their eccentricities~\citep{BarOr2018}.

One can follow a similar calculation to obtain the expression of the coherence
time, $\Tc^{\rt}$, defined in Eq.~\eqref{eq:Tct_def}
and characterizing the diffusive regime of the orientation's random walk.
It follows the power law distribution
\begin{equation}
\Tc^{\rt} (a , e) = \frac{P(a)}{\sqrt{N(\!<\!a)}} \frac{M_\bullet}{\sqrt{\langle m^2 \rangle}} \, \frac{f_{T} (e)}{\sqrt{\pi g(\gamma)}},
\label{res_Tct_pwl}
\end{equation}
where the dimensionless eccentricity function ${ f_{T} (e)}$
is given in Eq.~\eqref{def_T0_fT_pwl}
and illustrated in Fig.~\ref{fig:PowerLawf}.
In that same figure, we note that for a thermal eccentricity distribution
and a cusp's power index ${ 1 \leq \gamma \leq 2 }$, one can assume that
${ f_{\Gamma^{2}}^{1/2} (e) \, f_{T} (e) \simeq 0.4 }$, which leads
to the torque time and the coherence time following the approximate relation
\begin{equation}
\Tc^{\rt} (a , e) \simeq \frac{0.4}{\sqrt{1 - e^{2}}} \, \frac{1}{\Gamma (a , e)} .
\label{approx_Tct}
\end{equation}
We note that this simple relation allows for an even simpler generation
of samples of random walks in orientations as given by the toy model from Eq.~\eqref{eq:dipole_noise},
as one only has to estimate the test particle's torque time ${ 1 / \Gamma (a , e) }$,
as the associated coherence time, ${ \Tc^{\rt} (a , e) }$, follows immediately.

\section{Conclusion}
\label{sec:Conclusion}

In the present work, we illustrated how one can describe quantitatively the statistical properties of the stochastic evolutions of star's orientations in galactic nuclei during the process of \VRR\@. The main difficulty of the present derivation lies in the system being fundamentally degenerate, i.e. having a vanishing mean field Hamiltonian, ${ H = 0 }$.
This system is also non-Markovian, i.e.\ correlated in time, as well as non-ergodic, i.e.\ time- and ensemble-averages cannot be interverted.

Placing ourselves in the limit of an isotropic distribution of stars, we
circumvented some of these difficulties in Section~\ref{sec:CorrelNoise} by
assuming that the statistical properties of the noise fluctuations can be
derived from estimates of the derivatives of their correlation function at the
initial time. The main result was obtained in Eq.~\eqref{Gaussian_Cbath} which
provided us with a self-consistent ansatz for the statistical properties of
the time dependence of the correlation of the fluctuations generated jointly by
the system's $N$ particles. In Section~\ref{sec:TestParticle}, we used
this result to describe the random walk of a test particle's orientation embedded in this
stochastic system, recovering both the ballistic and diffusive regimes.  The
main result was obtained in Eq.~\eqref{Ct_dipole}, which yields quantitative
predictions for the statistical properties of that random walk.
The key tools used at that stage were the Magnus series to solve the linear
matrix evolution equation for the test particle, the independence hypothesis to
separate the statistics of the background noise from that of the test
particle's random walk, and the cumulant theorem to estimate ensemble averages.
We also emphasized how non-ergodic effects (associated with the constraint of
total energy conservation) should be accounted for to allow for reliable long
timescales predictions.  Throughout the text, all the predictions were compared
with detailed effective $N$-body simulations offering a quantitative agreement.
In Section~\ref{sec:SelfConsistency}, we highlighted the self-consistency
existing between the spontaneous fluctuations in the system
and the associated random walks in orientations.
Finally, in Section~\ref{sec:PowerLaw}, we presented a first application
of this framework to estimate the timescales of \VRR\ in a stellar cusp
similar to the one of SgrA*.

The present paper is only a first step towards a complete theory of \VRR\@, and
we list below some possible tracks for future developments.  In the current
derivation, we relied extensively on the isotropic assumption, and as such
neglected any effects associated with anisotropic clustering in
orientation~\citep{Szolgyen2018}.  For binaries, the exact statistical
properties of the \VRR\ random walk in orientation can lead to enhanced rates
of mergers~\citep{Hamers2018}, hence the importance for quantitative
predictions for the properties of these random walks, as obtained in
Eq.~\eqref{Ct_dipole}. Building upon Section~\ref{sec:TestParticle}, one could
also investigate how a substructure like a disk stochastically
dissolves~\citep{Kocsis2011}. This asks for a detailed accounting of the correlations in the
potential fluctuations of a given realization, to characterize how stars with
similar initial orientations or similar parameters get slowly separated.
Finally, here we focused our interest on systems dominated by a central mass.
Provided one updates accordingly the interaction coupling coefficients,
${ \mJ_{\ell} [ \bK , \bKp ] }$, similar investigations could be pursued
in the context of spherical globular clusters~\citep{Meiron2018}.

\acknowledgments
We thank Christophe Pichon and Scott Tremaine for remarks on an earlier version
of this manuscript.  JBF acknowledges support from Program number HST-HF2-51374
which was provided by NASA through a grant from the Space Telescope Science
Institute, which is operated by the Association of Universities for Research in
Astronomy, Incorporated, under NASA contract NAS5--26555.  BB is supported by
membership from Martin A. and Helen Chooljian at the Institute for Advanced
Study.

\appendix

\section{The coupling coefficients}
\label{sec:CouplingCoefficients}

Similarly to Eq.~{(10)} in~\cite{Kocsis2015}, we define the coupling coefficients\footnote{With this convention, the definition of $\mJ_{ij\ell}$ from Eq.~{(10)} of~\cite{Kocsis2015} is recovered by ${ \mJ_{ij\ell} = L [\bK_{i}] \,\mJ_{\ell} [\bK_{i} , \bK_{j}] \, (2 \ell + 1) / (4 \pi) }$.}
\begin{equation}
\label{def_mJ}
\mJ_{\ell} \big[ \bK_{i} , \bK_{j} \big] = \frac{G m_{i} m_{j}}{\aout} \frac{1}{L [\bK_{i}]} \, s_{\ell} \big[ \alpha , \ein , \eout \big],
\end{equation}
where ${ L [\bK] = m \sqrt{G \mBH a (1 - e^{2})} }$ is the magnitude of the
angular momentum.
We also introduced ``$\mathrm{out}$'' (resp. ``$\mathrm{in}$'') as the index $i$ or $j$ with the larger (resp. smaller) semi-major axis, and defined accordingly the ratio ${ \alpha = \ain / \aout \leq 1 }$.
With these notations, the dimensionless coefficients ${ s_{\ell} [\alpha , \ein , \eout] }$ are given by
\begin{equation}
s_{\ell} \big[ \alpha , \ein , \eout \big] = \frac{4 \pi \, P_{\ell}^{2} (0)}{2 \ell + 1} \, \frac{1}{\alpha} \,
\frac{1}{\pi^{2}} \!\! \int_{0}^{\pi} \!\! \rd \phi
\!\! \int_{0}^{\pi} \!\! \rd \phip \, \frac{\big( \text{min} \big[ \alpha (1 - \ein
  \cos (\phi)) , (1 - \eout \cos (\phip)) \big] \big)^{\ell + 1}}{\big(
  \text{max} \big[ \alpha (1 - \ein \cos (\phi)) , (1 - \eout \cos (\phip)) \big] \big)^{\ell}} ,
\label{def_os}
\end{equation}
with ${ P_{\ell} (u) }$ the usual Legendre polynomials. Because they are independent of the details of the considered system, the coefficients ${ s_{\ell} [\alpha , \ein , \eout] }$ can be precomputed on a grid to hasten the numerical evaluation of ${ \mJ_{\ell} [\bK_{i} , \bK_{j}] }$.
For our fiducial simulations, these coefficients were pre-computed on a linear ${3D}$ grid in ${ (\alpha , \ein , \eout) }$ consisting of $200^{3}$ elements, with ${ 10^{-2} \leq \alpha \leq 1 }$ and ${ 0 \leq \ein , \eout \leq 0.99 }$.
We refer to Fig.~{1} in~\cite{Kocsis2015} for an illustration of the behavior of these coefficients.

\section{The Elsasser coefficients}
\label{sec:Elsasser}

In this Appendix, we follow~\cite{James1973,Ivers2008} and detail some of the properties of the Elsasser coefficients.
We emphasize that we work with real spherical harmonics, hence the need for some identities to be adapted.

The real spherical harmonics are defined with the convention
\begin{equation}
Y_{\ell m} (u , \phi) = 
\begin{cases}
\displaystyle \sqrt{2} \, K_{\ell}^{|m|} \, P_{\ell}^{|m|} (u) \, \sin (|m| \phi) & \text{if} \; m < 0 ,
\\
\displaystyle K_{\ell}^{0} \, P_{\ell}^{0} (u) & \text{if} \; m = 0 ,
\\
\displaystyle \sqrt{2} \, K_{\ell}^{m} \, P_{\ell}^{m} (u) \, \cos (m \phi) & \text{if} \; m > 0 ,
\end{cases}
\label{def_ReYlm}
\end{equation}
with ${ P_{\ell}^{m} (u) }$ the usual associated Legendre polynomials, and the coefficients ${ K_{\ell}^{m} = \big[ \tfrac{2 \ell + 1}{4 \pi} \frac{(\ell - m)!}{(\ell + m)!} \big]^{1/2} }$. With this convention, the spherical harmonics follow the normalization ${ \! \int \! \rd \bL \, Y_{\ell m} Y_{\ellp \mpp} = \delta_{\ell}^{\ellp} \delta_{m}^{\mpp} }$.

Following Eq.~{(5.5)} of~\cite{Ivers2008}, the real Elsasser coefficients, as defined in Eq.~\eqref{definition_Elsasser_coefficients}, can be decomposed as
\begin{equation}
E_{\alpha \gamma \delta} = E^{L}_{\ell_{\alpha} \ell_{\gamma} \ell_{\delta}} \, E^{M}_{\alpha \gamma \delta} ,
\label{short_Elsasser}
\end{equation}
where $E^{L}_{\ell_{\alpha} \ell_{\gamma} \ell_{\delta}}$ only depends on ${ (\ell_{\alpha} , \ell_{\gamma} , \ell_{\delta}) }$, while ${ E^{M}_{\alpha \gamma \delta} }$ also depends on ${ (m_{\alpha} , m_{\gamma} , m_{\delta}) }$. The $m$-independent coefficients read
\begin{equation}
E^{L}_{\ell_{\alpha} \ell_{\gamma} \ell_{\delta}} = \Lambda_{\ell_{\alpha} \ell_{\gamma} \ell_{\delta}} \, \Delta_{\ell_{\alpha} \ell_{\gamma} \ell_{\delta}}
\begin{pmatrix}
\ell_{\alpha} + 1 & \ell_{\gamma} + 1 & \ell_{\delta} + 1
\\
0 & 0 & 0
\end{pmatrix} ,
\label{expression_EL}
\end{equation}
where we introduced the Wigner ${3j}$-symbols~\citep{Arfken2005}, and defined
\begin{align}
\Lambda_{\ell_{\alpha} \ell_{\gamma} \ell_{\delta}} & \, = \sqrt{\frac{(2 \ell_{\alpha} + 1) (2 \ell_{\gamma} + 1) (2 \ell_{\delta} + 1)}{4 \pi}} ,
\nonumber
\\
\Delta_{\ell_{\alpha} \ell_{\gamma} \ell_{\delta}} & \, = \sqrt{\frac{(\ell_{\alpha} + \ell_{\gamma} + \ell_{\delta} + 2) (\ell_{\alpha} + \ell_{\gamma} + \ell_{\delta} + 4)}{4 (\ell_{\alpha} + \ell_{\gamma} + \ell_{\delta} + 3)}} \, \sqrt{(\ell_{\alpha} + \ell_{\gamma} - \ell_{\delta} + 1) (\ell_{\delta} + \ell_{\alpha} - \ell_{\gamma} + 1) (\ell_{\gamma} + \ell_{\delta} - \ell_{\alpha} + 1)} .
\label{def_Lambda_Delta}
\end{align}
The coefficients $E^{M}_{\alpha \gamma \delta}$ are given by
\begin{equation}
E^{M}_{\alpha \gamma \delta} = \sum_{\mathclap{\veps_{\alpha},\veps_{\gamma},\veps_{\delta}=\pm1}} \, \text{Im} \big[ K_{\alpha \gamma \delta}^{\veps_{\alpha} \veps_{\gamma} \veps_{\delta}} \big] \,
\begin{pmatrix}
\ell_{\alpha} & \ell_{\gamma} & \ell_{\delta}
\\
\veps_{\alpha} m_{\alpha} & \veps_{\gamma} m_{\gamma} & \veps_{\delta} m_{\delta}
\end{pmatrix} ,
\label{expression_EM}
\end{equation}
where the tensor ${ K_{\alpha \gamma \delta}^{\veps_{\alpha} \veps_{\gamma} \veps_{\delta}} }$ comes from the fact that we are considering real spherical harmonics, and is given by
\begin{equation}
K_{\alpha \gamma \delta}^{\veps_{\alpha} \veps_{\gamma} \veps_{\delta}} = \kappa_{m_{\alpha}}^{\veps_{\alpha}} \, \kappa_{m_{\gamma}}^{\veps_{\gamma}} \, \kappa_{m_{\delta}}^{\veps_{\delta}}
\;\;\; \text{ with } \;\;\;
\kappa_{m}^{+1} = 
\begin{cases}
\displaystyle \ri \, (-1)^{m}/\sqrt{2} & \text{if } m < 0
\\
\displaystyle 1/2 & \text{if } m = 0
\\
\displaystyle 1/\sqrt{2} & \text{if } m > 0
\end{cases}
\;\;\; \text{ and } \;\;\;
\kappa_{m}^{-1} =
\begin{cases}
\displaystyle - \ri / \sqrt{2} & \text{if } m < 0
\\
\displaystyle 1/2 & \text{if } m = 0
\\
\displaystyle (-1)^{m} / \sqrt{2} & \text{if } m > 0 .
\end{cases}
\label{definition_K_EM}
\end{equation}
The Elsasser coefficients satisfy various exclusion rules~\citep{James1973}. In particular, for $E_{\alpha \gamma \delta}$ to be non-zero, one has to satisfy
\begin{align}
& \, \text{\{C1\}: }\; |m_{\alpha}| \leq \ell_{\alpha}; \;\; |m_{\gamma}| \leq \ell_{\gamma}; \;\; |m_{\delta}| \leq \ell_{\delta} ,
\nonumber
\\
& \, \text{\{C2\}: }\; \ell_{\alpha} + \ell_{\gamma} + \ell_{\delta} \text{ is odd},
\nonumber
\\
& \, \text{\{C3\}: }\; |\ell_{\alpha} - \ell_{\gamma}| < \ell_{\delta} < \ell_{\alpha} + \ell_{\gamma} \text{ (strict triangular inequality)},
\nonumber
\\
&\, \text{\{C4\}: }\; \text{all pairs } (\ell_{\alpha} , m_{\alpha}), (\ell_{\gamma} , m_{\gamma}), (\ell_{\delta} , m_{\delta}) \text{ are different}.
\label{exclusion_Elsasser}
\end{align}
These coefficients also follow the symmetry relations ${ E_{\alpha \delta \gamma} = E_{\gamma \alpha \delta} = - E_{\alpha \gamma \delta} }$.

Finally, following~\cite{QTAM}, the Elsasser coefficients satisfy various contraction identities. In particular, throughout the derivations, we will rely on
\begin{align}
& \, \sum_{\mathclap{m_{\gamma} , m_{\delta}}} \, E^{M}_{\alpha \gamma \delta} \, E^{M}_{\beta \gamma \delta} = \delta_{\alpha}^{\beta} \, \frac{1}{2 \ell_{\alpha} + 1}
\; \text{ for } \; |\ell_{\alpha} - \ell_{\beta}| < \ell_{\gamma/\delta} < |\ell_{\alpha} + \ell_{\beta}|
\; \text{ and } \; \ell_{\alpha} + \ell_{\beta} + \ell_{\gamma/\delta} \; \text{odd,} 
\nonumber
\\
& \, \sum_{\ell_{\delta}} \frac{1}{2 \ell_{\alpha} + 1} \big( E^{L}_{\ell_{\alpha} \ell_{\gamma} \ell_{\delta}} \big)^{2} = A_{\ell_{\alpha}} \, B_{\ell_{\gamma}} 
\; \text{ with } \;
A_{\ell} = \ell (\ell + 1) , \; B_{\ell} = \frac{\ell (\ell + 1) (2 \ell + 1)}{8 \pi} .
\label{sum_Elsasser}
\end{align}

\section{$N$-body simulations}
\label{sec:Nbody}

In this Appendix, we briefly detail the effective $N$-body simulations to which our analytical results are compared.
To simulate a system of $N$ interacting particles, the starting point is the evolution Eq.~\eqref{good_EOM},
that can be used for each of the $N$ particles.
In that form, we note that the velocity vector, ${ \rd \bL_{i} / \rd t }$, is expressed only as a function of the current location of the particle, ${ \bL_{i} (t) }$, and the instantaneous particle's magnetizations, ${ M_{\ell m} (\bK_{i} , t) }$.
There are $N$ such evolution equations, but because the magnetizations vary from one particle to another, their computation has to be made once per timestep and particle.
As a result, the overall complexity of advancing the particles for one timestep scales like ${ \mO (N^{2} \ell_{\rm max}^{2}) }$, with $\ell_{\rm max}$ the maximum harmonic number considered in the pairwise interaction\footnote{
We note the similarities between the evolution Eq.~\eqref{good_EOM}
and the evolution equation for the \HMF\ model~\citep{AntoniRuffo1995}.
The main differences are that (i) the present Hamiltonian has no kinetic term,
(ii) the magnetizations depend on two harmonic indices ${ (\ell , m) }$ (versus one for the \HMF\@ model),
(iii) the magnetizations are not global but vary from one particle
to another because of the parameters $\bK$.
As a result, the complexity of the integration of the \HMF\ model scales like ${ \mO (N \ell_{\rm max}) }$,
versus ${ \mO (N^{2} \ell_{\rm max}^{2}) }$ for the present model.
}.
In our approach, the motion of the particles is integrated by computing the magnetizations, while in the implementation presented in~\cite{Kocsis2015}, particles are moved forward by solving successively pairwise interactions, an approach symplectic by design.

Our $N$-body implementation proceeds then by (i) computing efficiently the spherical harmonics (and the vector ones) at the location of the particles, (ii) computing the magnetizations in Eq.~\eqref{def_Mlm}, (iii) computing the velocity fields in Eq.~\eqref{good_EOM}, (iv) advancing all the particles' orientation for one timestep. The real spherical harmonics are computed using a reccurence relation for the renormalized associated Legendre polynomials~\citep[see Eq.~{(6.7.9)} in][]{Press2007}, and using the second-order recurrence relation ${ \cos (m \phi) = 2 \cos (\phi) \cos ((m-1) \phi) - \cos ((m-2) \phi) }$ (similarly for ${ \sin (m \phi) }$) for the azimuthal component. To compute the real vector spherical harmonics, we follow the same recurrence as in Appendix~{(B.2)} of~\cite{Mignard2012}, adapted to the renormalized associated Legendre polynomials. Once all the velocity vectors ${ \rd \bL_{i} / \rd t }$ are computed, particles are advanced for a timestep $h$, using a fourth-order Runge-Kutta integrator~\citep[see Eq.~{(17.1.3)}][]{Press2007}.

All the derivations presented in the main text are illustrated with comparisons with this direct $N$-body approach.
We consider a system composed of ${ N = 10^{3} }$ stars, and assume that the particles' conserved quantities ${ \bK_{i} = (m_{i} , a_{i} , e_{i}) }$ satisfy ${ m = \mmin }$, ${ \amin \leq a \leq \amax }$, and ${ \emin \leq e \leq \emax }$.
Our units are chosen so that ${ \mmin = \amin = G = 1 }$, and we pick ${ \amax/\amin = 100 }$, ${ \emin = 0 }$, and ${ \emax = 0.3 }$.
These parameters are drawn independently one from another, according to \PDFs\ proportional to ${ (\deltaD (m - \mmin) , a^{1/2} , e) }$, which corresponds to a single-mass population in a harmonic profile with a thermal distribution of small eccentricities.
The stars' initial orientations are drawn uniformly on the sphere, and the interactions are truncated at ${ \ell_{\rm max} = 50 }$.
The timestep of the simulation is the same for all particles and is determined at the start of each realization. To do so, we compute the torque exerted on every particle at the initial time, ${ \tau_{i} = | \rd \bL_{i} / \rd t | = 1/ t_{\tau}^{i} }$, and define ${ t_{\tau}^{i} }$ as the associated torque time. The integration timestep is then fixed initially to ${ h = 10^{-2} \!\times\! \text{Min}_{i} [t_{\tau}^{i}] }$. With such choices, integrating the system for one timestep takes approximately ${1 \mathrm{s}}$ on a single core, and simulations are carried out for ${ 2 \!\times\! 10^{5} }$ timesteps.

\section{Computing the derivatives of the noise correlation}
\label{DerivativesNoise}

\subsection{Computing ensemble averages}
\label{sec:EnsembleAverage}

We are generically interested in computing ensemble averages at the initial time of the form ${ \savEA{\vphi_{\alpha} (\bK , 0) \, \vphi_{\beta} (\bKp , 0) \, ... } }$.
Such averages can be carried out explicitly by noting that at the initial time, the $N$ particles are drawn independently one from another, both for their orientations and their parameters.
Following our isotropic assumption, their orientation is drawn uniformly on the sphere, according to the \PDF\ ${ f (\bL) = 1/(4\pi) }$, while we assume that their parameter $\bK$ is drawn according to a \PDF\@, ${ g (\bK) }$, normalized so that ${ \!\int\! \rd \bK \, g(\bK) = 1 }$.

To illustrate the gist of these calculations, let us consider the case ${ \savEA{\vphi_{\alpha} (\bK , 0) \, \vphi_{\beta} (\bKp , 0)} }$. As they do not contribute to the dynamics, we never need to consider the harmonics ${ (\ell , m) = (0 , 0) }$, so that ${ \!\int\! \rd \bL f (\bL) Y_{\alpha/\beta} (\bL) = 0 }$.
Owing to the particle independence at the initial time and following the definition from Eq.~\eqref{def_vphi_alpha}, we can write
\begin{align}
\avEA{\vphi_{\alpha} (\bK , 0) \, \vphi_{\beta} (\bKp , 0)} = & \, \!\! \int \!\! \rd \bL_{1} \rd \bK_{1} \, ... \, \rd \bL_{N} \rd \bK_{N} \, f (\bL_{1}) \, g (\bK_{1}) \, ... \, f (\bL_{N}) \, g (\bK_{N})
\nonumber
\\
& \, \times \sum_{i , j} \deltaD(\bK - \bK_{i}) \, \deltaD (\bKp - \bK_{j}) \, Y_{\alpha} (\bL_{i}) \, Y_{\beta} (\bL_{j})
\nonumber
\\
= & \, N \avc{\vphi_{\alpha} (\bK , 0) \, \vphi_{\beta} (\bKp , 0)} . 
\label{average_two}
\end{align}
where non-zero terms only come from ${ i = j }$, and we introduced the connected average as
\begin{align}
\avc{\vphi_{\alpha} (\bK , 0) \, \vphi_{\beta} (\bKp , 0)} & \, = \!\! \int \!\! \rd \bL_{1} \rd \bK_{1} \, f (\bL_{1}) \, g (\bK_{1}) \, \deltaD (\bK - \bK_{1}) \, \deltaD (\bKp - \bK_{1}) \, Y_{\alpha} (\bL_{1}) \, Y_{\beta} (\bL_{1}) 
\nonumber
\\
& \, = \delta_{\alpha}^{\beta} \, \deltaD (\bK - \bKp) \, \frac{g (\bK)}{4 \pi} .
\label{def_connected_average}
\end{align}
When considering averages at the initial time involving more than two fields, we limit ourselves to the dominant contributions associated with pair couplings (i.e. the limit of Gaussian fields, Wick's theorem).
We can then write
\begin{equation}
\avEA{ \vphi_{1} (\bK_{1} , 0) \, ... \, \vphi_{n} (\bK_{n} , 0) } = 
\begin{cases}
\displaystyle N^{n/2} \big( \avc{\vphi_{1} \, \vphi_{2} } \, ... \, \avc{\vphi_{n-1} \, \vphi_{n} } + \text{perm.} \big) & \, \text{ if } $n$ \text{ even},
\\
\displaystyle 0 & \, \text{ if } $n$ \text{ odd},
\end{cases}
\label{average_full}
\end{equation}
where ``$\text{perm}$'' browses all the possible pair decompositions without repetitions, and averages involving an odd number of fields are neglected.

\subsection{Initial values of the correlation function}
\label{sec:CorrelInit}

Following the method just described, we may now estimate the value
and the second derivative of the correlation function at the initial time,
as introduced in Eq.~\eqref{def_C}.

For the value at the initial time, we can write
\begin{align}
C_{\alpha \beta} (\bK , \bK' , 0) & \, = \avEA{\vphi_{\alpha} (\bK , 0) \, \vphi_{\beta} (\bK' , 0)}
\nonumber
\\
& \, = \delta_{\alpha}^{\beta} \, \deltaD (\bK - \bK') \, n (\bK) .
\label{res_C0}
\end{align}
where we followed Eq.~\eqref{average_two} to compute the last average, and
introduced ${ n (\bK) = g (\bK) \, N / (4 \pi) }$ as the \DF\ of the stars' parameters
satisfying the normalization ${ \!\int\! \rd \bL \rd \bK \, n (\bK) \!=\! N }$.
We emphasize that to compute Eq.~\eqref{res_C0}, we relied on the assumption of an
isotropic distribution of particles on the sphere, which led to the Kronecker
coefficients $\delta_{\alpha}^{\beta}$ w.r.t.\ the harmonic coefficients.

The ensemble average expectation for the first derivative at the initial time
reads
${ \partial C_{\alpha \beta} / \partial t \!=\! \savEA{\dot{\vphi}_{\alpha} \, \vphi_{\beta}}
  \!\sim\! \savEA{\vphi_{\gamma} \, \vphi_{\delta} \vphi_{\beta}} \!=\! 0 }$,
where we used the quadratic evolution Eq.~\eqref{EOM_vphi_II} once. As it involves
an odd number of fields, this correlation is equal to zero,
as imposed by Eq.~\eqref{average_full}.

Let us now turn to the computation of the ensemble-average expectation for the second derivative of the correlation at the initial time.
We write
\begin{align}
\frac{\partial^{2}}{\partial t^{2}} C_{\alpha \beta} (\bK , \bK' , t) \bigg|_{t = 0} \!\! & \, = \avEA{\ddot{\vphi}_{\alpha} (\bK , 0) \, \vphi_{\beta} (\bK' , 0)}
\nonumber
\\
 & \, = - \avEA{\dot{\vphi}_{\alpha} (\bK , 0) \, \dot{\vphi}_{\beta} (\bK' , 0)}
\nonumber
\\
& \, = - \!\! \int \!\! \rd \bK_{1} \rd \bK_{2} \, \mJ_{\ell_{\gamma}} \big[ \bK , \bK_{1} \big] \mJ_{\ell_{\delta}} \big[ \bK' , \bK_{2} \big] E_{\alpha \gamma \lambda} E_{\beta \delta \rho}
\nonumber
\\
& \;\;\;\; \times \avEA{\vphi_{\gamma} (\bK_{1} , 0) \, \vphi_{\lambda} (\bK , 0) \, \vphi_{\delta} (\bK_{2} , 0) \, \vphi_{\rho} (\bK' , 0)} .
\label{calc_C2_bath}
\end{align}
where we injected the evolution Eq.~\eqref{EOM_vphi_II} twice.
As shown in Eq.~\eqref{average_full}, in the limit of Gaussian fluctuations, the average term can be computed by keeping only averages of pairs.
As ${ E_{\alpha \gamma \gamma} = 0 }$, only two of the possible couplings remain, namely ${ \avEA{\gamma \delta} \, \avEA{\lambda \rho} }$ and ${ \avEA{\gamma \rho} \, \avEA{\delta \lambda} }$, which leads to
\begin{align}
\frac{\partial^{2}}{\partial t^{2}} C_{\alpha \beta} (\bK , \bK'  , t) \bigg|_{t = 0} \!\! & \, = - E_{\alpha \gamma \delta} \, E_{\beta \gamma \delta} \, \big[ \Gamma_{\gamma} (\bK , \bK') - \Lambda_{\gamma \delta} (\bK , \bK') \big] , 
\label{calc_C2_II_bath}
\end{align}
where we used ${ E_{\beta \delta \gamma} = - E_{\beta \gamma \delta} }$, and introduced
\begin{align}
& \, \Gamma_{\gamma} (\bK , \bK') \!=\! \deltaD (\bK \!-\! \bK') \, n (\bK) \!\! \int \!\! \rd \bKpp \, n (\bKpp) \, \mJ_{\ell_{\gamma}}^{2} \big[ \bK , \bKpp \big] ,
\nonumber
\\
& \, \Lambda_{\gamma \delta} (\bK , \bK') \!=\! n (\bK) \, n (\bK') \, \mJ_{\ell_{\gamma}} \big[ \bK , \bK' \big] \, \mJ_{\ell_{\delta}} \big[ \bK' , \bK \big] .
\label{def_Lambda_Gamma}
\end{align}
Following Eq.~\eqref{sum_Elsasser}, one can now perform the sums over $m_{\gamma}$ and $m_{\delta}$ in Eq.~\eqref{calc_C2_II_bath}.
The \VRR\ interactions being limited to even harmonic numbers $\ell$, we may impose at this stage that $\ell_{\alpha}$ is even. Glancing back at the constraint~{\{C2\}} from Eq.~\eqref{exclusion_Elsasser}, we note that ${ \ell_{\alpha} \!+\! \ell_{\gamma} \!+\! \ell_{\delta} }$ has to be odd, so that the term ${ \Lambda_{\gamma \delta} (\bK , \bK') }$ never contributes to Eq.~\eqref{calc_C2_II_bath}. For $\ell_{\alpha}$ even, Eq.~\eqref{calc_C2_II_bath} becomes
\begin{align}
\frac{\partial^{2}}{\partial t^{2}} C_{\alpha \beta} (\bK , \bK' , t) \bigg|_{t = 0} \!\! = & \, - \delta_{\alpha}^{\beta} \, A_{\ell_{\alpha}} \, \deltaD (\bK - \bK') \, n (\bK) \, \Gamma^{2} (\bK) ,
\label{res_C2}
\end{align}
where the sum over $\ell_{\delta}$ was performed following Eq.~\eqref{sum_Elsasser},
and the decay rate ${ \Gamma^{2} (\bK) }$ is given by Eq.~\eqref{Gamma_bath}.

\section{Computing the properties of the random walk}
\label{sec:CalcRandomWalk}

In this Appendix, we compute the two averages appearing in Eq.~\eqref{calc_Ct}.
Assuming that the test particle is initially uniformly distributed on the sphere,
one straightforwardly has
\begin{equation}
\avLt{\vphit_{\alpha} (0) \, \vphit_{\beta} (0)} \! = \!\! \int \!\! \frac{\rd \bLt}{4 \pi} \, Y_{\alpha} (\bLt) \, Y_{\beta} (\bLt) = \frac{\delta_{\alpha}^{\beta}}{4 \pi} .
\label{init_average_bath}
\end{equation}
To compute the time average of ${ \re^{\Omega (t)} }$, we rely on the cumulant theorem,
\begin{equation}
\avT{\re^{\veps \, \Omega}} = \sum_{n = 0}^{\infty} \frac{\veps^{n}}{n!} \, \mu_{n} = \exp \!\bigg[ \sum_{n = 1}^{\infty} \frac{\veps^{n}}{n!} \kappa_{n} \bigg] ,
\label{cumulant_theorem}
\end{equation}
where ${ \mu_{n} = \avT{\Omega^{n}} }$ are the moment matrices, and
$\kappa_{n}$ are the cumulant matrices, with the first two ones given by
${ \kappa_{1} = \mu_{1} }$ and ${ \kappa_{2} = \mu_{2} - \mu_{1}^{2} }$.
Here, we compute the time average of ${ \re^{\Omega (t)} }$ by keeping only terms that are
at most second order in $\Qt$, so that only ${ \savT{\Omega_{1} (t)} \!\propto\! \Qt }$
and ${ \savT{\Omega_{1}^2 (t)} \!\propto\! \savT{\Omega_{2} (t)} \!\propto\! {(\Qt)}^2 }$
contribute. We note that since ${ \savT{\vphi_\alpha(\bK,t)} \!=\! 0 }$ and, from
stationarity, ${ \savT{\Qt(t) \, \Qt(\tp)} \!=\! \savT{\Qt(\tp) \, \Qt(t)} }$, both
$\savT{\Omega_{1} (t)} $ and $\savT{\Omega_2 (t)}$ vanish. As a result, at the
order considered here, only the second cumulant
$\kappa_2 = \savT{\Omega_{1}^2 (t)}$ is non-zero, and from the cumulant theorem
we obtain
\begin{equation}
{\Big[ \navT{\re^{\Omega (t)}} \Big]}_{\alpha \beta} = {\Big[ \re^{ \frac{1}{2} \savT{\Omega_1^{2} (t)}}  \Big]}_{\alpha \beta}
\label{cumulant_average}
\end{equation}
Gathering Eqs.~\eqref{init_average_bath} and~\eqref{cumulant_average} together,
we can write the test particle's correlation function as
\begin{equation}
C_{\alpha\beta}^{\rt , \rr} (t) = \frac{1}{4 \pi} \, {\Big[ \re^{ \frac{1}{2} \savT{\Omega_1^{2} (t)}}  \Big]}_{\alpha \beta} .
\label{rewrite_together_correlation}
\end{equation}

Using Eq.~\eqref{series_O}, we can write
\begin{align}
\label{calc_cumulant}
  & \, {\bigg[ \bavT{\Omega_1^2 (t)} \bigg]}_{\alpha
    \beta} \!\! = 
    \!\! \int_{0}^{t} \!\! \rd s_{1} \!\! \int_{0}^{t} \!\! \rd s_{2} \!\! \int
    \!\! \rd \bK_{1}  \!\! \int \!\! \rd \bK_{2} \, E_{\alpha \gamma \delta} E_{\delta \lambda \beta}
    \nonumber
  \\
  & \, \times \, \mJ_{\ell_{\gamma}} \big[ \bKt , \bK_{1} \big] \, 
    \mJ_{\ell_{\lambda}} \big[ \bKt , \bK_{2} \big] \,
    \avT{\vphi_{\gamma} (\bK_{1} , s_{1}) \, 
    \vphi_{\lambda} (\bK_{2} , s_{2})} .
\end{align}
To pursue the calculation further, we may now use our ansatz for the
time-dependence of the correlation of the noise fluctuations, as obtained in
Eq.~\eqref{Gaussian_Cbath}. Using the sum identities from
Eq.~\eqref{sum_Elsasser}, one gets
\begin{align}
  \label{calc_cumulant_II}
  {\bigg[ \bavT{\Omega_1^2 (t)} \bigg]}_{\alpha \beta} \!\! 
  & =  - \delta_{\alpha}^{\beta} \, A_{\ell_{\alpha}} \sum_{\ell} B_{\ell}
    \!\! \int \!\! \rd \bK \, 
    \mJ_{\ell}^2 \big[ \bKt , \bK \big] \chi_{\ell}^{\rr} (\bK , t) ,
\end{align}
where ${ \chi_{\ell}^{\rr} (\bK , t) }$ is a double time integral of the noise correlation
\begin{align}
  \label{def_omC}
  \chi_{\ell}^{\rr} (\bK , t) 
  & = \!\! \int_{0}^{t} \!\! \rd t_{1} \!\!
    \int_{0}^{t} \!\! \rd t_{2} \, C_{\ell}^{\rr}
    (\bK , t_{1} - t_{2}) .
    \nonumber \\
  & = n_{\ell}^{\rr} (\bK) \, \frac{2\Tc^2(\bK)}{A_{\ell}}\chi\bigg[
  \frac{t\sqrt{A_{\ell}/2}}{\Tc(\bK)}\bigg],
\end{align}
with the dimensionless function ${ \chi (\tau) }$ defined in Eq.~\eqref{eq:def_chi}.
Since the matrix $\savT{\Omega_1^2 (t)}$ is
diagonal, one can straightforwardly compute its exponential, as required by
Eq.~\eqref{cumulant_average}.  This allows us to finally recast the correlation of the
test particle's random motion from Eq.~\eqref{rewrite_together_correlation} under the form of Eq.~\eqref{final_Ct}.

\section{Computing the variance of the noise amplitude}
\label{sec:NoiseVariance}

In this Appendix, we compute the ensemble-averaged variance of the amplitude of the density fluctuations, ${ C_{\alpha\beta}^{\rr} (\bK_{\alpha} , \bK_{\beta} , 0) }$, as introduced in Eq.~\eqref{def_C_real}.
Our goal is to compute an expression of the form
\begin{equation}
\avEA{C_{\alpha\beta}^{\rr} (\bK_{\alpha} , \bK_{\beta} , 0) \, C_{\gamma \delta}^{\rr} (\bK_{\gamma} , \bK_{\delta} , 0)} = \frac{1}{T^{2}} \!\! \int_{0}^{T} \!\! \rd t \!\! \int_{0}^{T} \!\! \rd \tp \, \avEA{ \vphi_{\alpha} (\bK_{\alpha} , t) \, \vphi_{\beta} (\bK_{\beta} , t) \, \vphi_{\gamma} (\bK_{\gamma} , \tp) \, \vphi_{\delta} (\bK_{\delta} , \tp) } ,
\label{square_C}
\end{equation}
Because only even harmonics contribute to the interactions, we can limit ourselves to ${ 2 \leq \ell_{\alpha} , \ell_{\gamma} }$ even.
As estimated in Eq.~\eqref{def_oC}, we know that for ${ |t - \tp| \gg \Tc }$, the location of the background particles at time $\tp$ can be considered to be decorrelated from their locations at time $t$, up to the requirement of satisfying the system's global conservation constraints.
It is fundamental to account for these global conservation constraints, as they introduce non-ergodic effects, preventing us from interverting time- and ensemble-averages.
Provided that these constraints are satisfied, in the two-dimensional time integral from Eq.~\eqref{square_C}, one can note that the particles are uncorrelated between $t$ and $\tp$ on a surface of size ${ (T - \Tc)^{2} }$, while they are correlated on a surface of size ${ T \Tc }$.
As a result, as long as ${ T \gg \Tc }$ and as long as the conservation constraints are satisfied, particles can be considered as uncorrelated between time $t$ and $\tp$, and therefore distributed uniformly over the sphere at these two times.

Let us now detail how one may carry out the average from Eq.~\eqref{square_C}, in the presence of these constraints.
At time $t$, the state of the system is fully characterized by the set of all fields ${ \bvphi \!=\! \{ \vphi_{\alpha} (\bK_{\alpha} , t) \} }$, and similarly, at time $\tp$, the state of the system is fully characterized by ${ \bvphip \!=\! \{ \vphip_{\alpha} (\bK_{\alpha} , t) \} }$.
We may then use $\bvphi$ and $\bvphip$ as the random variables over which averages are carried out.
Following Eq.~\eqref{C0_bath}, we have
\begin{equation}
\avEA{\vphi_{\alpha} (\bK_{\alpha} , 0)} = 0 \;\;\; \text{(for ${ \ell_{\alpha} \neq 0 }$)} ; \;\;\; \avEA{\vphi_{\alpha} (\bK_{\alpha} , 0) \, \vphi_{\beta} (\bK_{\beta} , 0)} = \delta_{\alpha}^{\beta} \, \deltaD (\bK_{\alpha} - \bK_{\beta}) \, n (\bK_{\alpha}) ,
\label{Gaussian_statistics}
\end{equation}
and similarly for $\bvphip$.
Placing ourselves within the Gaussian limit, we may then treat $\bvphi$ (resp. $\bvphip$) as uncorrelated Gaussian random fields, that follow a Gaussian \PDF\ ${ F (\bvphi) }$ (resp. ${ F (\bvphip) }$), with a covariance following from Eq.~\eqref{Gaussian_statistics}.

As emphasized above, the two fields $\bvphi$ and $\bvphip$ still remain correlated one with another through global constraints. To shorten the notation, let us temporarily note these constraints as ${ \bT = \bT (\bvphi) }$. In Eq.~\eqref{square_C}, the average must then be carried out according to the joint \PDF\ ${ F (\bvphi , \bvphip) = F (\bvphi) \, F (\bvphip \, | \, \bT (\bvphi)) }$.
The conditional \PDF\ of $\bvphip$ given the constraint ${ \bT (\bvphi) }$ follows from Bayes theorem, and reads
\begin{equation}
F (\bvphip \, | \, \bT (\bvphi)) = \frac{F (\bvphip , \bT (\bvphi))}{F_{\bT} (\bT (\bvphi))} = \frac{F (\bvphip) \, \deltaD (\bT (\bvphi) - \bT (\bvphip))}{F_{\bT} (\bT (\bvphi))} ,
\label{conditional_PDF_vphip}
\end{equation}
with ${ F_{\bT} (\bT) }$ the \PDF\ of the constraints $\bT$.
Therefore, we can write
\begin{equation}
F (\bvphi , \bvphip) = \frac{F (\bvphi) \, F (\bvphip) \, \deltaD (\bT (\bvphi) - \bT(\bvphip))}{F_{\bT} (\bT (\bvphi))} .
\label{rewrite_joint_PDF}
\end{equation}
In that view, Eq.~\eqref{square_C} can be recast as
\begin{equation}
\avEA{C^{\rr}_{\alpha\beta} (\bK_{\alpha} , \bK_{\beta} , 0) \, C^{\rr}_{\gamma\delta} (\bK_{\gamma} , \bK_{\delta} , 0)} = \!\! \int \!\! \rd \bT \rd \bTp \, \frac{\deltaD (\bT - \bTp)}{F_{\bT} (\bT)} \, \dF_{\alpha \beta} (\bK_{\alpha} , \bK_{\beta} , \bT) \, \dF_{\gamma \delta} (\bK_{\gamma} , \bK_{\delta} , \bTp) ,
\label{square_C_joint}
\end{equation}
where we introduced
\begin{align}
\dF_{\alpha\beta} (\bK_{\alpha} , \bK_{\beta} , \bT) & \, \equiv \!\! \int \!\! \rd \bvphi \, F (\bvphi) \, \vphi_{\alpha} (\bK_{\alpha}) \, \vphi_{\beta} (\bK_{\beta}) \, \deltaD (\bT (\bvphi) - \bT)
\nonumber
\\
& \, = \avEA{\vphi_{\alpha} (\bK_{\alpha}) \, \vphi_{\beta} (\bK_{\beta}) \, \deltaD (\bT (\bvphi) - \bT)} ,
\label{def_dF_full}
\end{align}
with ${ \savEA{ \, \cdot \,} }$ standing for the ensemble average where the fields $\vphi$ are drawn according to the Gaussian statistics of ${ F(\bvphi) }$.
Conveniently, in that form, Eq.~\eqref{square_C_joint} allows us to carry out independently the averages over $\bvphi$ and $\bvphip$.

To proceed further, let us now detail the global conservation constraints that have to be satisfied throughout the system's evolution.
There are three such constraints, namely the conservation of each particle's individual parameters ${ (\bT_{0}) }$, the conservation of the system's total angular momentum ${ (\bT_{1}) }$, and the conservation of the system's total energy ${ (\bT_{2}) }$. Luckily, these can all be expressed as simple functions of the fields $\bvphi$. They read
\begin{align}
& \, \text{$\bK$-conservation ($\bT_{0}$):} && \hspace{-0.8cm} \vphi_{(0 , 0)} (\bK , t) = \text{cst.} ,
\nonumber
\\
& \, \text{Angular momentum conservation ($\bT_{1}$):} && \hspace{-0.8cm} \frac{1}{N} \!\! \int \!\! \rd \bK \, L \big[ \bK \big] \, \vphi_{(1, m)} (\bK , t) = \text{cst.} \;\; \text{for} \; -1 \leq m \leq 1 ,
\nonumber
\\
& \, \text{Energy conservation ($\bT_{2}$):} && \hspace{-0.8cm} E = \frac{1}{2 N} \sum_{\alpha} \!\! \int \!\! \rd \bK \rd \bKp \, H_{\ell_{\alpha}} \big[ \bK , \bKp \big] \, \vphi_{\alpha} (\bK ,t) \, \vphi_{\alpha} (\bKp , t) = \text{cst.} ,
\label{def_constraint}
\end{align}
with ${ L [\bK] \!=\! m \sqrt{G \mBH a (1 - e^{2})} }$ the norm of the angular momentum and ${ H_{\ell} [\bK , \bKp] = L [\bK] \, \mJ_{\ell} [\bK , \bKp] }$.
We also note the prefactor ${ 1 / (2 N) }$ in the definition of the energy that was introduced for later convenience.
At this stage, it is important to note that each of these constraints involve different harmonics of the Gaussian random fields, namely ${ \ell = 0 }$ for the conservation of $\bK$, ${ \ell = 1 }$ for the angular momentum, and ${ 2 \leq \ell }$ even for the energy.
In the limit of Gaussian random fields, this implies that only the energy constraint contributes to a non-zero variance in Eq.~\eqref{square_C_joint}, as we will now argue.

Since only even harmonics contribute to the interactions, we can restrict ourselves to ${ 2 \leq \ell_{\alpha} }$ even when computing ${ \dF_{\alpha\beta} (\bK_{\alpha} , \bK_{\beta} , \bT) }$.
If ${ \ell_{\beta} = 0 , 1 }$, Eq.~\eqref{def_dF_full} can be rewritten as
\begin{equation}
\dF_{\alpha\beta} (\bK_{\alpha} , \bK_{\beta} , \bT) = \avEA{ \vphi_{\beta} (\bK_{\beta}) \, \delta_{D} (\bT_{0} (\bvphi) - \bT_{0}) \, \deltaD (\bT_{1} (\bvphi) - \bT_{1}) } \, \avEA{ \vphi_{\alpha} (\bK_{\alpha}) \, \deltaD ( \bT_{2} (\bvphi) - \bT_{2}) } ,
\label{rewrite_dF_one}
\end{equation}
where we used the Gaussian assumption, so that fields with different harmonics are uncorrelated. Because the energy is quadratic in the fields, and because the Gaussian \PDF\@, ${ F (\bvphi_{\ell \geq 2}) }$, is an even function of the fields, the last bracket in Eq.~\eqref{rewrite_dF_one} is equal to zero.
As a result, we can assume that ${ \ell_{\beta} \geq 2 }$.
In that case, Eq.~\eqref{def_dF_full} becomes
\begin{align}
\dF_{\alpha\beta} (\bK_{\alpha} , \bK_{\beta} , \bT) & \, = \avEA{\deltaD (\bT_{0} (\bvphi) - \bT_{0})} \, \avEA{\deltaD (\bT_{1} (\bvphi) - \bT_{1})} \, \avEA{\vphi_{\alpha} (\bK_{\alpha}) \, \vphi_{\beta} (\bK_{\beta}) \, \deltaD (\bT_{2} (\bvphi) - \bT_{2})}
\nonumber
\\
& \, = F_{\bT_{0}} (\bT_{0}) \, F_{\bT_{1}} (\bT_{1}) \, \dF_{\alpha\beta} (\bK_{\alpha} , \bK_{\beta} , E) ,
\label{rewrite_dF_two}
\end{align}
where we introduced
\begin{equation}
\dF_{\alpha\beta} (\bK_{\alpha} , \bK_{\beta} , E) \equiv \avEA{\vphi_{\alpha} (\bK_{\alpha}) \, \vphi_{\beta} (\bK_{\beta}) \, \deltaD (E (\bvphi) - E)} .
\label{def_dF}
\end{equation}
As a result, for ${ \ell_{\alpha} , \ell_{\beta} , \ell_{\gamma} , \ell_{\delta} \geq 2}$, this allows us to rewrite the needed correlation from Eq.~\eqref{square_C_joint} as
\begin{equation}
\avEA{C_{\alpha\beta}^{\rr} (\bK_{\alpha} , \bK_{\beta} , 0) \, C_{\gamma\delta}^{\rr} (\bK_{\gamma} , \bK_{\delta} , 0)} = \!\! \int \!\! \rd E \rd \Ep \, \frac{\deltaD (E - \Ep)}{F_{E} (E)} \, \dF_{\alpha\beta} (\bK_{\alpha} , \bK_{\beta} , E) \, \dF_{\gamma\delta} (\bK_{\gamma} , \bK_{\delta} , \Ep) ,
\label{square_C_joint_E}
\end{equation}
where we got rid of all occurences of the constraints $\bT_{0}$ and $\bT_{1}$, using the fact that their \PDFs\ satisfy ${ \!\int\! \rd \bT_{0} \, F_{\bT_{0}} \!=\! \!\int\! \rd \bT_{1} F_{\bT_{1}} \!=\! 1 }$.
As a conclusion, in the limit of Gaussian random fields, only the constraint of total energy conservation contributes to the non-ergodic properties of the system.
This is an important result of this calculation.

Let us now compute the ensemble average appearing in Eq.~\eqref{def_dF}.
In the present Gaussian limit, we can rely on Novikov theorem~\citep{Novikov1965} to compute it.\footnote{We do not repeat here the general theory of Novikov theorem~\citep{Novikov1965}. We refer to~\cite{Hanggi1978} for non-Gaussian noises, to~\cite{Garcia1999} for spatially extended noises, and to~\cite{FouvryBarOr2018} for an example of application in stellar dynamics.}
One gets
\begin{equation}
\dF_{\alpha\beta} (\bK_{\alpha} , \bK_{\beta} , E) = \sum_{\mu} \!\! \int \!\! \rd \bK_{\mu} \, \avEA{\vphi_{\alpha} (\bK_{\alpha}) \, \vphi_{\mu} (\bK_{\mu})} \, \avEA{ \frac{\delta}{\delta \vphi_{\mu} (\bK_{\mu})} \bigg[ \vphi_{\beta} (\bK_{\beta}) \, \deltaD (E (\bvphi) - E) \bigg] } ,
\label{Novikov_dF}
\end{equation}
where the first cumulant is absent because ${ \ell_{\alpha} \geq 2 }$, so that ${ \savEA{\vphi_{\alpha} (\bK_{\alpha})} = 0 }$, and only the second cumulant remains as the fields are assumed to be Gaussian.
In Eq.~\eqref{Novikov_dF}, the sum (resp. integral) over $\mu$ (resp. ${ \rd \bK_{\mu} }$) runs over all the fields. 
The functional gradient appearing in the last term can be computed as
\begin{equation}
\frac{\delta }{\delta \vphi_{\mu} (\bK_{\mu})} \bigg[ \vphi_{\beta} (\bK_{\beta}) \, \deltaD (E (\bvphi) - E) \bigg] = \delta_{\mu}^{\beta} \, \deltaD (\bK_{\mu} - \bK_{\beta}) \, \deltaD (E (\bvphi) - E) - \vphi_{\beta} (\bK_{\beta}) \, \frac{\partial }{\partial E} \bigg[ \frac{\delta E (\bvphi)}{\delta \vphi_{\mu} (\bK_{\mu})} \, \deltaD (E (\bvphi) - E) \bigg] ,
\label{calc_func_grad}
\end{equation}
where we used the fundamental relation ${ \delta \vphi_{\beta} (\bK_{\beta}) / \delta \vphi_{\mu} (\bK_{\mu}) = \delta_{\mu}^{\beta} \deltaD (\bK_{\beta} - \bK_{\mu}) }$. Glancing back at the definition of the energy in Eq.~\eqref{def_constraint}, we can also write
\begin{equation}
\frac{\delta E (\bvphi)}{\delta \vphi_{\mu} (\bK_{\mu})} = \frac{1}{N} \!\! \int \!\! \rd \bK \, H_{\ell_{\mu}} \big[ \bK_{\mu} , \bK \big] \, \vphi_{\mu} (\bK) .
\label{func_grad_E}
\end{equation}
Injecting these results into Eq.~\eqref{Novikov_dF} and using the Gaussian statistics from Eq.~\eqref{Gaussian_statistics}, we obtain a self-consistent integro-differential equation for ${ \dF_{\alpha \beta} (\bK_{\alpha} , \bK_{\beta} , E) }$, namely
\begin{equation}
\dF_{\alpha\beta} (\bK_{\alpha} , \bK_{\beta} , E) = \delta_{\alpha}^{\beta} \, \deltaD (\bK_{\alpha} - \bK_{\beta}) \, n (\bK_{\alpha}) \, F_{E} (E) - \frac{1}{N} \, n (\bK_{\alpha}) \!\! \int \!\! \rd \bK \, H_{\ell_{\alpha}} \big[ \bK_{\alpha} , \bK \big] \, \frac{\partial }{\partial E} \dF_{\alpha \beta} (\bK , \bK_{\beta} , E) ,
\label{Fredholm_equation}
\end{equation}
where we used that ${ \savEA{\deltaD (E (\bvphi) - E)} = F_{E} (E) }$, by definition.

Progress can now be made by accounting perturbatively for the total energy constraint.
As such, we introduce the small parameter $\veps$, make the substitution ${ H_{\ell} \to \veps H_{\ell}  }$ in Eq.~\eqref{Fredholm_equation}, and consider the expansion
\begin{equation}
\dF = \dF^{(0)} + \veps \, \dF^{(1)} + \veps^{2} \, \dF^{(2)} + ... 
\label{expansion_dF}
\end{equation}
We can then inject this expansion in Eq.~\eqref{Fredholm_equation} and match the orders in $\veps$.
The first three terms are obtained as
\begin{align}
\dF_{\alpha\beta}^{(0)} (\bK_{\alpha} , \bK_{\beta} , E) & \, = \delta_{\alpha}^{\beta} \, \deltaD (\bK_{\alpha} - \bK_{\beta}) \, n (\bK_{\alpha}) \, F_{E} (E) ,
\nonumber
\\
\dF_{\alpha\beta}^{(1)} (\bK_{\alpha} , \bK_{\beta} , E) & \, = - \frac{1}{N} \delta_{\alpha}^{\beta} \, n (\bK_{\alpha}) \, n (\bK_{\beta}) \, \frac{\partial F_{E} (E)}{\partial E} \, H_{\ell_{\alpha}} \big[ \bK_{\alpha} , \bK_{\beta} \big] ,
\nonumber
\\
\dF_{\alpha\beta}^{(2)} (\bK_{\alpha} , \bK_{\beta} , E) & \, = \frac{1}{N^{2}} \delta_{\alpha}^{\beta} \, n (\bK_{\alpha}) \, n (\bK_{\beta}) \, \frac{\partial^{2} F_{E} (E)}{\partial E^{2}} \!\! \int \!\! \rd \bK \, H_{\ell_{\alpha}} \big[ \bK , \bK_{\beta} \big] \, H_{\ell_{\alpha}} \big[ \bK_{\alpha} , \bK \big] \,n (\bK) .
\label{dF_first_terms}
\end{align}

Owing to these first terms, we can now return to the computation of the variance from Eq.~\eqref{square_C_joint_E}.
Keeping only terms at most second order in $\veps$, this reads
\begin{align}
& \hspace{-5.4cm} \savEA{C_{\alpha\beta}^{\rr} \, C_{\gamma\delta}^{\rr}} = \savEA{C_{\alpha\beta}^{\rr} \, C_{\gamma\delta}^{\rr}}^{(0)} + \veps \, \savEA{C_{\alpha\beta}^{\rr} \, C_{\gamma\delta}^{\rr}}^{(1)} + \veps^{2} \, \savEA{C_{\alpha\beta}^{\rr} \, C_{\gamma\delta}^{\rr}}^{(2)} + ...
\nonumber
\\
 =  \!\! \int \!\! \rd E \rd \Ep \, \frac{\deltaD (E - \Ep)}{F_{E} (E)} \, \bigg[ & \, \dF_{\alpha\beta}^{(0)} (E) \, \dF_{\gamma \delta}^{(0)} (\Ep)
\nonumber
\\
+ \,  \veps \, \bigg\{ & \, \dF_{\alpha\beta}^{(0)} (E) \, \dF_{\gamma\delta}^{(1)} (\Ep) + \dF_{\alpha\beta}^{(1)} (E) \, \dF_{\gamma\delta}^{(0)} (\Ep) \bigg\}
\nonumber
\\
+ \, \veps^{2} \, \bigg\{ & \, \dF_{\alpha\beta}^{(0)} (E) \, \dF_{\gamma\delta}^{(2)} (\Ep) + \dF_{\alpha\beta}^{(1)} (E) \, \dF_{\gamma\delta}^{(1)} (\Ep) + \dF_{\alpha\beta}^{(2)} (E) \, \dF_{\gamma\delta}^{(0)} (\Ep) \bigg\} \bigg] ,
\label{injection_variance}
\end{align}
where, for simplicity, we did not repeat the arguments ${ (\bK_{\alpha} , \bK_{\beta}) }$ and ${ (\bK_{\gamma} , \bK_{\delta}) }$.
The zeroth-order term is straightforward to compute, and gives
\begin{align}
\savEA{C_{\alpha\beta}^{\rr} (\bK_{\alpha} , \bK_{\beta} , 0) \, C_{\gamma\delta}^{\rr} (\bK_{\gamma} , \bK_{\delta} , 0)}^{(0)} & = \delta_{\alpha}^{\beta} \, \delta_{\gamma}^{\delta} \, \deltaD (\bK_{\alpha} - \bK_{\beta}) \, \deltaD (\bK_{\gamma} - \bK_{\delta}) \, n (\bK_{\alpha}) \, n (\bK_{\gamma})
\nonumber
\\
& \, = C_{\alpha\beta} (\bK_{\alpha} , \bK_{\beta} , 0) \, C_{\gamma\delta} (\bK_{\gamma} , \bK_{\delta} , 0) ,
\label{varC_0}
\end{align}
using ${ \!\int\! \rd E \, F_{E} (E) = 1 }$.
It is straighforward to show that the first-order term satisfies
\begin{equation}
\savEA{C_{\alpha\beta}^{\rr} (\bK_{\alpha} , \bK_{\beta} , 0) \, C_{\gamma\delta}^{\rr} (\bK_{\gamma} , \bK_{\delta} , 0)}^{(1)} = 0 ,
\label{varC_1}
\end{equation}
as the energies integrals vanish.
Using a similar argument, one finds that terms of the form ${ \dF^{(0)} \, \dF^{(2)} }$ do not contribute to the second-order term in Eq.~\eqref{injection_variance}.
Keeping only the non-zero contribution coming from ${ \dF_{\alpha\beta}^{(1)} (E) \, \dF_{\gamma\delta}^{(1)} (\Ep) }$, we get
\begin{equation}
\savEA{C_{\alpha\beta}^{\rr} (\bK_{\alpha} , \bK_{\beta} , 0) \, C_{\gamma\delta}^{\rr} (\bK_{\gamma} , \bK_{\delta} , 0)}^{(2)} = \frac{1}{N^{2}} \, \delta_{\alpha}^{\beta} \, \delta_{\gamma}^{\delta} \, n (\bK_{\alpha}) \, n (\bK_{\beta}) \, n (\bK_{\gamma}) \, n (\bK_{\delta}) \, \frac{H_{\ell_{\alpha}} \big[ \bK_{\alpha} , \bK_{\beta} \big] \, H_{\ell_{\gamma}} \big[ \bK_{\gamma} , \bK_{\delta} \big]}{(\Delta E)^{2}} ,
\label{varC_2}
\end{equation}
where ${ \Delta E }$ is obtained after straightforward manipulations of the energy integrals, and reads
\begin{equation}
\frac{1}{(\Delta E)^{2}} \equiv \!\! \int \!\! \rd E \, \frac{1}{F_{E} (E)} \, \bigg( \frac{\partial F_{E} (E)}{\partial E} \bigg)^{2} . 
\label{def_DeltaE}
\end{equation}
It is important to note that ${ \Delta E }$ is a single number that depends only on the total energy \PDF\@, ${ F_{E} (E) }$, and therefore only on the considered \DF\, ${ n (\bK) }$.
In Appendix~\ref{sec:EnergyVariance}, we detail how the needed integral from Eq.~\eqref{def_DeltaE} can be estimated.
Equation~\eqref{varC_2} is an important result of this Appendix, as it characterizes the variance (over different realizations) of the noise fluctuations' amplitude arising from the constraint of total energy conservation.

Following Eq.~\eqref{def_oC_m}, we can get the variance of ${ n_{\ell}^{r} (\bK) }$.
It reads
\begin{equation}
\avEA{n_{\ell}^{\rr} (\bK) \, n_{\ellp}^{\rr} (\bKp)} = n (\bK) \, n (\bKp) \, \bigg\{ 1 + M_{\ell} (\bK) \, M_{\ellp} (\bKp) \bigg\} ,
\label{var_oC}
\end{equation}
where we introduced the dimensionless function ${ M_{\ell} (\bK) }$ as
\begin{equation}
M_{\ell} (\bK) = \!\! \int \!\! \rd \bKp \, \frac{n (\bKp)}{N} \, \frac{H_{\ell} \big[ \bK , \bKp \big]}{\Delta E} .
\label{def_M}
\end{equation}
The final step of this Appendix is to compute the variance of $\Gamma_{\rt}^{2}$, as defined in Eq.~\eqref{def_kappa}.
We get
\begin{equation}
\savEA{(\Delta \Gamma_{\rt}^{2})^{2}} = \bigg\{ \sum_{\ell} B_{\ell} \!\! \int \!\! \rd \bK \, n (\bK) \, \mJ_{\ell}^{2} \big[ \bK_{\rt} , \bK \big] \, M_{\ell} (\bK) \bigg\}^{2} ,
\label{var_Ct}
\end{equation}
with ${ \Delta \Gamma_{\rt}^{2} \!=\! \Gamma_{\rt}^{2} \!-\! \Gamma^{2} }$.
Equation~\eqref{var_Ct} is the final result of this section.
It expresses the variance (over realizations) of ${ \Gamma_{\rt}^{2} }$, the amplitude of the random walk of a given test star.
It is important to note that this since ${ (\Gamma^{2})^{2} }$ and ${ \savEA{(\Delta \Gamma_{\rt}^{2})^{2}} }$ have the same scaling with $N$, the present variance effect does not vanish as the number of particles gets larger.

\begin{figure}
\begin{center}
\includegraphics[width=0.48\textwidth]{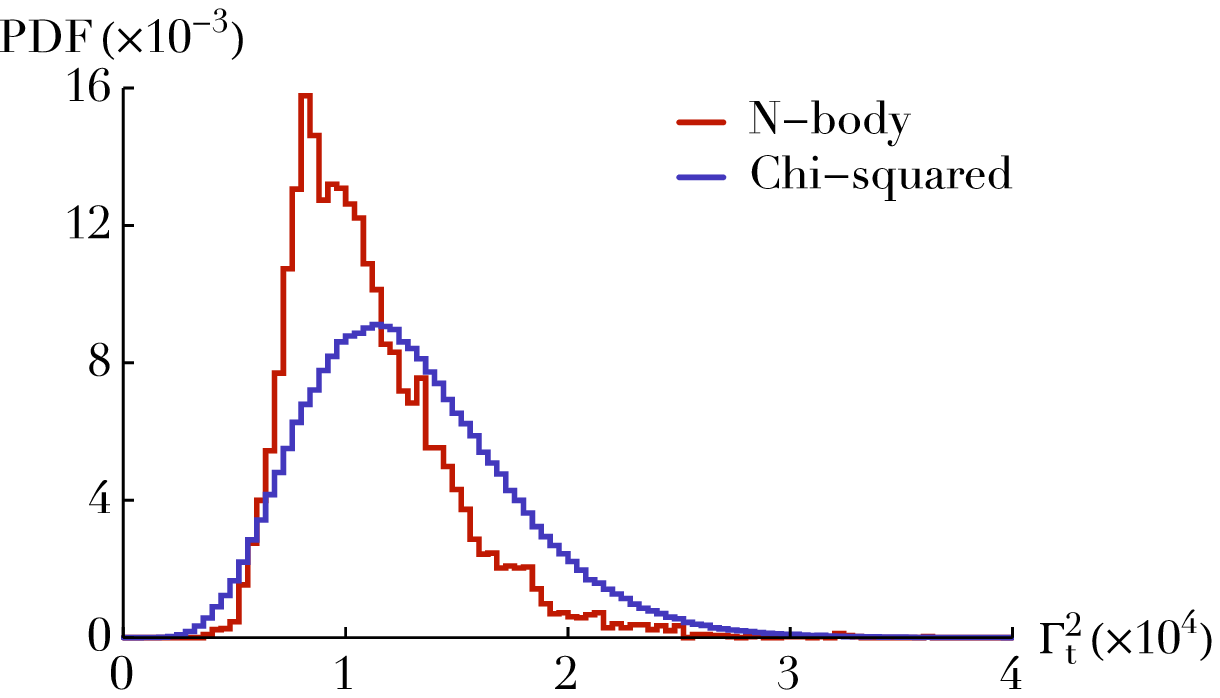}
\caption{Illustration of the variation in $\Gamma_{\rt}^{2}$ for the test particles falling in the same window ${ W_{\rt} (\bKt) }$ as in Fig.~\ref{fig:RandomWalk}.
The red histogram is the distribution of $\Gamma_{\rt , \num}^{2}$ measured over ${1000}$ realizations.
This distribution is characterized by ${ \savEA{\Gamma_{\rt}^{2}}_{\num} \simeq  1.1 \!\times\! 10^{-4} }$ and ${ \kappa_{\num} \!=\! \savEA{\Gamma_{\rt}^{2}}_{\num}^{2} / \savEA{(\Delta \Gamma_{\rt}^{2})^{2}}_{\num} \!\simeq\! 9.3 }$.
The blue histogram has been obtained via a resampling of $\Gamma_{\rt}^{2}$ following chi-squared distributions with means and variances predicted in Eqs.~\eqref{def_kappa} and~\eqref{var_Ct}.
This distribution is characterized by ${ \savEA{\Gamma_{\rt}^{2}}_{\pred} \simeq 1.3 \!\times\! 10^{-4} }$ and ${ \kappa_{\pred} \simeq 7.7 }$.
\label{fig:HistogramC0t}}
\end{center}
\end{figure}

As it will be needed to obtain the prediction of Fig.~\ref{fig:RandomWalk}, let us illustrate the effect associated with this non-zero variance of ${ \Gamma_{\rt}^{2} }$ in our fiducial simulations.
We consider the same window, ${ W_{\rt} (\bKt) }$, as in Eq.~\eqref{def_Wt}. For each test particle falling in that window, we measure the correlation function ${ \savT{\vphi_{\alpha}^{\rt} (t) \, \vphi_{\alpha}^{\rt} (0)} }$ (e.g.\ for ${ \ell_{\alpha} = 1 }$).
The second-order time derivative at ${ t = 0 }$ of this correlation function is directly proportional to ${ \Gamma_{\rt , \num}^{2} }$ (see Eq.~\eqref{Ct_dipole}), that we can therefore measure numerically.
In Fig.~\ref{fig:HistogramC0t}, we represent the distribution of these numerically measured initial values, ${ \Gamma_{\rt , \num}^{2} }$, and illustrate how these amplitudes vary from realizations to realizations.

To capture this variance effect seen in Fig.~\ref{fig:HistogramC0t}, we may use our estimation of the variance of $\Gamma_{\rt}^{2}$ obtained in Eq.~\eqref{var_Ct}.
To do so, for every test particle falling in the window, we compute ${ (\Gamma^{2} , \savEA{(\Delta \Gamma_{\rt}^{2})^{2}}) }$, following Eqs.~\eqref{kappa_mean} and~\eqref{var_Ct}.
Having determined this mean and variance, one can draw a sample of ${ \Gamma_{\rt}^{2} }$ according to a \PDF\ that shares the same first two cumulants.
Similarly to Eq.~\eqref{def_E2}, we assume that we can draw this sample according to a chi-squared \PDF\ with these imposed mean and variance.
The result of this procedure is illustrated in Fig.~\ref{fig:HistogramC0t}.
This figure illustrates how the present calculations are able to capture most of the features
associated with the non-vanishing variance of the test particles'
individual amplitudes ${\Gamma_{\rt}^{2} }$.

\section{Computing the variance of the energy distribution}
\label{sec:EnergyVariance}

In this Appendix, we briefly detail how one can estimate
the energy spread ${ \Delta E }$ introduced in Eq.~\eqref{def_DeltaE}.
Our convention for the definition of the total energy is spelled out in Eq.~\eqref{def_constraint}.
Using the two-point statistics from Eq.~\eqref{def_connected_average}, the expected mean value of the energy reads
\begin{equation}
\avEA{E} = \frac{1}{2} \sum_{\ell} (2 \ell + 1) \, E_{\ell}^{(1)} ,
\label{mean_E}
\end{equation}
where we defined
\begin{equation}
E_{\ell}^{(1)} = \!\! \int \!\! \rd \bK \, H_{\ell} \big[ \bK , \bK \big] \, \frac{n (\bK)}{N} .
\label{def_E1}
\end{equation}

We can proceed similarly to compute the expectation for ${ \savEA{E^{2}} }$. This reads
\begin{align}
\savEA{E^{2}} = \frac{1}{4} \sum_{\alpha , \beta} \! \int \!\! \rd \bK_{\alpha} \bKp_{\alpha} \rd \bK_{\beta} \rd \bKp_{\beta} \, H_{\ell_{\alpha}} \big[ \bK_{\alpha} , \bKp_{\alpha} \big] \, H_{\ell_{\beta}} \big[ \bK_{\beta} , \bKp_{\beta} \big] \, \avEA{\vphi_{\alpha} (\bK_{\alpha}) \, \vphi_{\alpha} (\bKp_{\alpha}) \, \vphi_{\beta} (\bK_{\beta}) \, \vphi_{\beta} (\bKp_{\beta})} .
\label{var_E_calc}
\end{align}
To compute the average term appearing in the r.h.s.\@, we follow Eq.~\eqref{average_full}, placing ourselves in the limit of Gaussian random fields so that only connected averages involving two fields remain.
We get
\begin{equation}
\savEA{E^{2}} - \savEA{E}^{2} = \frac{1}{2} \sum_{\ell} (2 \ell + 1) \, E_{\ell}^{(2)} ,
\label{var_E}
\end{equation}
where we used the symmetry relation ${ H_{\ell} [\bK , \bKp] = H_{\ell} [\bKp , \bK] }$, and introduced
\begin{equation}
E_{\ell}^{(2)} = \!\! \int \!\! \rd \bK \rd \bKp \, H_{\ell}^{2} \big[ \bK , \bKp \big] \, \frac{n (\bK)}{N} \, \frac{n (\bKp)}{N} .
\label{def_E2}
\end{equation}

\begin{figure}
\begin{center}
\includegraphics[width=0.48\textwidth]{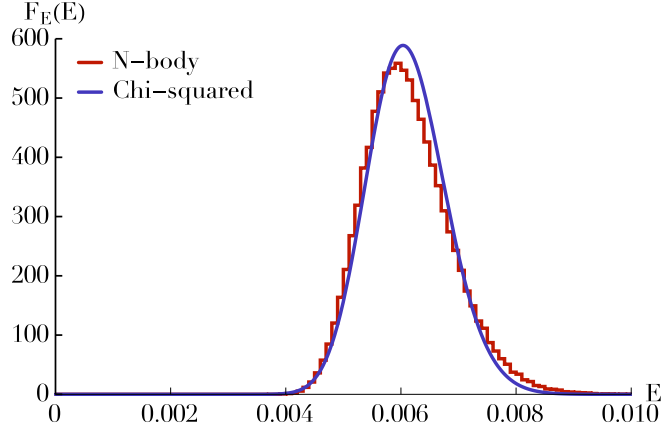}
\caption{Illustration of the statistical distribution of the system's total energy.
The red histogram has been estimated numerically by computing the initial energy of ${ 2 \!\times\! 10^{5} }$ realizations.
This histogram is characterized by ${ \savEA{E}_{\num} \simeq 6.1 \times\! 10^{-3} }$ and ${ \kappa_{\num} \simeq 65 }$.
The black line corresponds to the chi-squared \PDF\ prediction from Eq.~\eqref{PDF_Chi}, for which ${ \savEA{E}_{\mathrm{chi}} \simeq 6.1 \!\times\! 10^{-3} }$ and ${ \kappa_{\rm chi} \simeq 81 }$.
\label{fig:SamplingEnergy}}
\end{center}
\end{figure}

Having estimated the mean and the variance of the energy distribution, we may now return to the evaluation of the energy spread ${ \Delta E }$ introduced in Eq.~\eqref{def_DeltaE}.
As the energy is a quadratic function of Gaussian fields, we will assume that its follows a (scaled) chi-squared distribution of mean ${ \mu = \savEA{E} }$, and variance ${ \sigma^{2} = \savEA{E^{2}} - \savEA{E}^{2} }$.
The associated \PDF\ then follows
\begin{equation}
F_{E} (E) = \frac{1}{\mu} \frac{\kappa}{\Gamma (\kappa)} \, \bigg( \kappa \, \frac{E}{\mu} \bigg)^{\kappa - 1} \! \re^{- \kappa E / \mu} .
\label{PDF_Chi}
\end{equation}
with ${ \kappa = \mu^{2} / \sigma^{2} }$.
For our fiducial numerical system, we find ${ \savEA{E} \simeq 6.1 \!\times\! 10^{-3}  }$ and ${ \kappa \simeq 81  }$.
In Fig.~\ref{fig:SamplingEnergy}, we illustrate the statistical distribution of the system's energy, as well as the approximation from Eq.~\eqref{PDF_Chi}.
Finally, for a chi-squared \PDF\ as in Eq.~\eqref{PDF_Chi}, one can explicitly compute ${ \Delta E }$, as defined in Eq.~\eqref{def_DeltaE}, to get
\begin{equation}
\Delta E = \avEA{E} \frac{\sqrt{\kappa - 2}}{\kappa} .
\label{calc_DeltaE}
\end{equation}
We note that this integral is well-behaved only for ${ \kappa > 2 }$.
This is an artefact coming from the perturbative expansion introduced in Eq.~\eqref{expansion_dF}.
For our fiducial model, we find ${ \Delta E \simeq 6.7 \!\times\! 10^{-4}  }$.

\section{Computing averages over window}
\label{sec:WindowAverage}

In this Appendix, we briefly detail the procedures used in Figs.~\ref{fig:CorrelationOphi} and~\ref{fig:RandomWalk} to compare our analytical results with the fiducial numerical simulations.

\subsection{Correlation of the noise fluctuations}
\label{sec:WindowOphi}

Let us detail the method followed to obtain Fig.~\ref{fig:CorrelationOphi}
used to illustrate Eq.~\eqref{Gaussian_Cbath}.
One of the key insight from this equation is that to any particle (of parameter $\bK$), we can associate the pair ${ ( n (\bK) , \Tc (\bK)) }$ that characterizes the correlation properties of the density fluctuations generated by background particles with these parameters.
As a result, in order to consider only particles that have similar noise decorrelation properties, it is convenient to introduce, for every realization, the $\bK$-averaged fields ${ \ophi_{\alpha} (t) \!=\! \!\int\! \rd \bK \, W (\bK) \, \vphi_{\alpha} (\bK , t) }$, with ${ W (\bK) }$ a window function defined\footnote{As detailed in Appendix~\ref{sec:Nbody}, our fiducial simulations are single-mass, so that ${ n (\bK) \!\propto\! \deltaD (m -\mmin) \, g_{a} (a) \, g_{e} (e) }$. As a consequence, for the definition of the window function ${ W (\bK) }$ in Eq.~\eqref{def_W} to be meaningful, we do not account for the Dirac delta in mass present in ${ n (\bK) }$.} as
\begin{equation}
W (\bK) \!=\! 
\begin{cases}
\displaystyle 1 & \displaystyle \!\!\!\! \text{if } \, 1 \leq \frac{n (\bK)}{C_{\mathrm{min}}} , \frac{\Tc (\bK)}{\Tc^{\mathrm{min}}} \leq 1 + \veps_{W} ,
\\
\displaystyle 0 & \displaystyle \!\!\!\! \text{otherwise} .
\end{cases}
\label{def_W}
\end{equation}
with ${ (C_{\mathrm{\min}} , \Tc^{\mathrm{\min}}) }$ the typical amplitude and torque time considered, and ${ \veps_{W} }$ a small dimensionless parameter controlling the size of the window.
We then naturally have ${ \ophi_{\alpha} (t) \!=\! \sum_{i \in W} Y_{\alpha} (\bL_{i} (t)) }$, with the sum limited to the particles with ${ ( n (\bK) , \Tc (\bK)) }$ in the vicinity of ${ ( C_{\mathrm{min}} , \Tc^{\mathrm{min}}) }$.
Introducing ${ C_{\ell_{\alpha} , W}^{\rr} (t - \tp) \equiv \savT{\ophi_{\alpha} (t) \, \ophi_{\alpha} (\tp)} }$, Eq.~\eqref{Gaussian_Cbath} immediately gives
\begin{equation}
C_{\ell , W}^{\rr} (t) = \!\! \int \!\! \rd \bK \, W (\bK) \, C_{\ell}^{\rr} (\bK , t) ,
\label{WAveraged_Correlation_Bath}
\end{equation}
where the function ${ C_{\ell}^{\rr} (\bK , t) }$ follows the Gaussian ansatz from Eq.~\eqref{def_oC}.
When averaged over realizations, Eq.~\eqref{WAveraged_Correlation_Bath} can be approximated with the Gaussian dependence
\begin{equation}
C_{\ell , W}^{\rr} (t) \simeq C_{W} \, \re^{- \frac{A_{\ell}}{2} (t / T_{W})^{2}} ,
\label{Approximation_Correlation_Bath}
\end{equation}
where we introduced the amplitude ${ C_{W} }$ and torque time ${ T_{W} }$ as
\begin{equation}
C_{W} \!=\! \! \int \! \rd \bK \, W (\bK) \, n (\bK) ; \;\;\; \frac{C_{W}}{(T_{W})^{2}} \!=\! \!\int\! \rd \bK \, W (\bK) \frac{n (\bK)}{(\Tc (\bK))^{2}} .
\label{def_oCW_TW}
\end{equation}
Equation~\eqref{Approximation_Correlation_Bath} is the analytical Gaussian
prediction illustrated in Fig.~\ref{fig:CorrelationOphi}.

In Fig.~\ref{fig:CorrelationOphi}, we also present an updated prediction of the noise correlation
obtained by reinjecting the Gaussian prediction from Eq.~\eqref{eq:C_bath_exp}
into the self-consistency relation from Eq.~\eqref{self_relation}.
In that context, when averaged over the window, the prediction takes the form
\begin{equation}
C_{\ell , W} (t) \simeq C_{W} \exp \bigg\{ - \frac{A_{\ell}}{2} \!\! \int_{0}^{t} \!\! \rd t_{1} \!\! \int_{0}^{t} \!\! \rd t_{2} \, \Gamma_{W}^{2} \, \exp^{ - [ (t_{1} - t_{2}) / T_{W}^{\rt} ]^{2} } \bigg\} ,
\label{reinjected_prediction}
\end{equation}
with $C_{W}$ as in Eq.~\eqref{def_oCW_TW}, and
where the amplitude $\Gamma_{W}^{2}$ and coherence time ${ T_{W}^{\rt} }$ are given by
\begin{equation}
\Gamma_{W}^{2} = \avW{ \Gamma_{\rt}^{2} (\bK)} ; \;\;\; T_{W}^{\rt} = \avW{ T_{\rc}^{\rt} (\bK)} .
\label{def_CW_TW}
\end{equation}
In these equations, we introduced ${ \avW{ \, \cdot \, } }$ as the mean over the window ${ W (\bK) }$,
i.e.\ it is defined as
\begin{equation}
\avW{f (\bK)}= \frac{\! \int \! \rd \bK \, n (\bK) \, W (\bK) \, f (\bK)}{\! \int \! \rd \bK \, n (\bK) \, W (\bK)} .
\label{def_av_W}
\end{equation}

\subsection{Correlation of the random walks}
\label{sec:WindowRandomWalk}

Let us briefly detail the method followed to obtain Fig.~\ref{fig:RandomWalk}, used to illustrate the result from Eq.~\eqref{Ct_dipole}.
Following the independence hypothesis from Eq.~\eqref{calc_Ct}, we assume that for a given realization, each individual particle can effectively be treated as a test particle.
As such, we neglect the correlations existing between the background fluctuations and the random walk of that one particular particle.

One important insight from Eq.~\eqref{Ct_dipole} is that to any test particle (of parameter $\bKt$), we can associate the pair ${ (\Gamma^{2} (\bK_{\rt}) , \Tc^{\rt} (\bK_{t})) }$, that characterizes the correlation properties of its random walk in orientation.
Similarly to Eq.~\eqref{def_W}, in order to investigate these random walks, it is convenient to introduce the window function ${ W_{\rt} (\bKt) }$ as
\begin{equation}
W_{\rt} (\bKt) \!=\! 
\begin{cases}
\displaystyle 1 & \displaystyle \!\!\!\! \text{if } \, 1 \leq \frac{\Gamma^{2} (\bK_{\rt})}{\Gamma_{\mathrm{min}}^{2}} , \frac{\Tc^{\rt} (\bKt)}{T_{\rt}^{\mathrm{min}}} \leq 1 + \veps_{W_{\rt}} ,
\\
\displaystyle 0 & \displaystyle \!\!\!\! \text{otherwise} ,
\end{cases}
\label{def_Wt}
\end{equation}
where ${ (\Gamma_{\mathrm{min}}^{2} , T_{\rt}^{\mathrm{min}}) }$ are the typical amplitude and coherence time of the considered test particles, and $\veps_{W_{\rt}}$ is a dimensionless parameter controlling the size of the window. 
We may then define the window-averaged correlation function (that can easily be measured in the numerical simulations) as
\begin{equation}
C_{\ell_{\alpha} , W_{\rt}}^{\rt} (t) = 4 \pi \avEA{\avWt{ \vphi_{\alpha}^{\rt} (t) \, \vphi_{\alpha}^{\rt} (0)}} ,
\label{def_Ct_mean}
\end{equation}
where ${ \savWt{ \, \cdot \, } }$ stands for the mean over the test particles of a given realization falling in the window ${ W_{\rt} (\bKt) }$, similarly to Eq.~\eqref{def_av_W}.
Following Eq.~\eqref{Ct_dipole}, we also added a prefactor ${ 4 \pi }$ to ensure that this correlation is between $0$ and $1$.

If one does not account for the variance in ${ \Gamma_{\rt}^{2} }$ (see Eq.~\eqref{var_Ct}), a first (naive) prediction for Eq.~\eqref{def_Ct_mean} can be obtained from Eq.~\eqref{Ct_dipole} by restricting ourselves only to the ensemble-averaged mean prediction-s. This gives
\begin{equation}
C_{\ell , W_{\rt}}^{\rt} (t) \simeq \exp \bigg\{ - \frac{A_{\ell}}{2} \, \!\! \int_{0}^{t} \!\! \rd t_{1} \!\! \int_{0}^{t} \!\! \rd t_{2} \, \Gamma_{W_{\rt}}^{2} \, \re^{- [(t_{1} - t_{2}) / T_{W_{\rt}}^{\rt}]^{2}} \bigg\} ,
\label{Naive_Ct_mean}
\end{equation}
where the amplitude, ${ \Gamma_{W_{\rt}}^{2} }$, and coherence time, ${ T_{W_{\rt}} }$, are computed by direct averages over the particles falling in the window ${ W_{\rt} (\bKt) }$, so that
\begin{equation}
\Gamma_{W_{\rt}}^{2} = \avWt{\Gamma^{2} (\bK_{\rt})} ; \;\;\; T_{W_{\rt}}^{\rt} = \avWt{\Tc^{\rt} (\bK_{\rt})} . 
\label{Naive_Ct_other}
\end{equation}

One can improve the prediction from Eq.~\eqref{Naive_Ct_mean} by accounting for
the variance in $\Gamma_{\rt}^{2}$.
To do so, for every test particle falling in
the window, one can compute the mean expectation for the amplitude,
${ \savEA{\Gamma_{\rt}^{2} (\bK_{\rt})} = \Gamma^{2} (\bK_{\rt}) }$, and the associated variance,
${ \savEA{ (\Delta \Gamma_{\rt}^{2} (\bK_{\rt}) )^{2} } }$, as given by
Eqs.~\eqref{kappa_mean} and~\eqref{var_Ct}.  For this same particle, one can then
draw an effective value of ${ \Gamma_{\rt}^{2} }$, according to a
chosen \PDF\ with these prescribed mean and variance. This process is
illustrated in Fig.~\ref{fig:HistogramC0t}, where we used a chi-squared \PDF\@.
In that case, the prediction from Eq.~\eqref{Naive_Ct_mean} becomes
\begin{equation}
C_{\ell , W_{\rt}}^{\rt} (t) \simeq \avWt{ \exp \bigg\{ - \frac{A_{\ell}}{2} \!\! \int_{0}^{t} \!\! \rd t_{1} \!\! \int_{0}^{t} \!\! \rd t_{2} \, \Gamma_{\rt}^{2} \, \re^{- [ (t_{1} - t_{2})/T_{W_{\rt}}^{\rt} ]^{2}} \bigg\} } .
\label{variance_Ct_mean}
\end{equation}
Here, the amplitude and coherence time from Eq.~\eqref{Naive_Ct_other} become
\begin{equation}
\Gamma_{\rt}^{2} = \mathrm{PDF} \big[ \Gamma^{2} , {\savEA{(\Delta \Gamma_{\rt}^{2})^{2}}} \big] ; \;\;\; T_{W_{\rt}}^{\rt} = \Tc^{\rt} ,
\label{variance_Ct_other}
\end{equation}
where ${ \mathrm{PDF} [\mu , \sigma^{2}] }$ returns a sample from a chosen
\PDF\ of mean $\mu$ and variance $\sigma^{2}$ (which we chose to be a
chi-squared \PDF\ as in Fig.~\ref{fig:HistogramC0t}).  Both predictions from
Eqs.~\eqref{Naive_Ct_mean} and~\eqref{variance_Ct_mean} are illustrated in
Fig.~\ref{fig:RandomWalk}, for the particular harmonic ${ \ell = 1 }$.

\section{The case of a power law distribution}
\label{sec:CalcPowerLaw}

In this Appendix, we detail all the calculations presented in Section~\ref{sec:PowerLaw}
for an infinite power law stellar distribution around a \MBH\@.
We first note that the squared coupling coefficients from Eq.~\eqref{def_mJ}
can be rewritten as
\begin{equation}
\mJ_{\ell}^{2} \big[ \bK , \bKp \big] = \frac{G}{\mBH} \, \frac{1}{a (1 - e^{2})} \, \frac{\mpsq}{\aout^{2}} \, s_{\ell}^{2} \big[ \alpha , \ein , \eout \big] ,
\label{squared_coupling_pwl}
\end{equation}
where we recall that ``$\mathrm{out}$'' (resp. ``$\mathrm{in}$'') labels the star with the larger (resp. smaller) semi-major axis,
and we introduced the dimensionless ratio ${ \alpha = \ain / \aout }$.

\begin{figure}
\begin{center}
\includegraphics[width=0.55\textwidth]{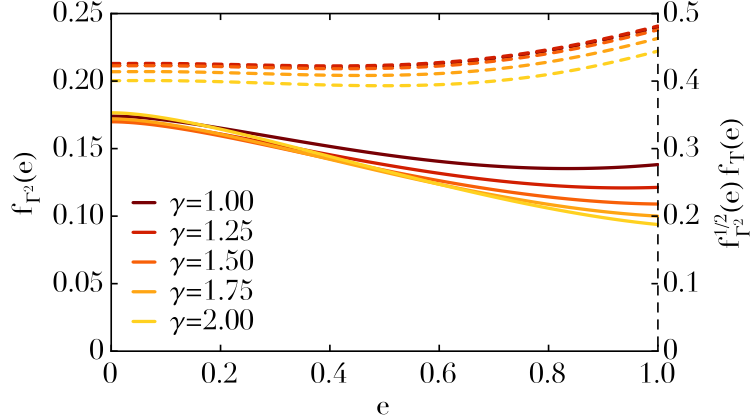}
\caption{
Illustration of the dimensionless eccentricity functions ${ f_{\Gamma^{2}} (e) }$ (left axis, full lines) and ${ f_{\Gamma^{2}}^{1/2} (e) \, f_{T} (e) }$ (right axis, dashed lines), for different cusp's profiles (through the power index $\gamma$) and assuming a thermal eccentricity distribution, ${ f_{e} (e) = 2e }$.
We note that ${ f_{\Gamma^{2}}^{1/2} (e) \, f_{T} (e) \simeq 0.4 }$
independently of $e$ and $\gamma$.
Calculations of the integrals over $\alpha$ were performed using the
same grid in $s_{\ell}$ as in Appendix~\ref{sec:CouplingCoefficients}.
\label{fig:PowerLawf}}
\end{center}
\end{figure}

Following Eq.~\eqref{Gamma_bath}, we can then compute the amplitude, $\Gamma^{2}$,
of the background fluctuations.
It reads
\begin{equation}
\Gamma^{2} (\bK) = \frac{1}{4 \pi} \frac{G \, \avmsq}{\mBH} \, \frac{1}{a (1 - e^{2})} \sum_{\ell} B_{\ell} \!\! \int \!\! \rd \app \rd \epp \, \frac{f_{e} (\epp) \, n_{a} (\app)}{\aout^{2}} \, s_{\ell}^{2} \big[ \tfrac{\ain}{\aout} , \ein , \eout \big] ,
\label{calc_Gamma_pwl}
\end{equation}
where we introduced the second moment of the mass distribution
${ \avmsq \!=\! \!\int\! \rd m f_{m} (m) \, m^{2} }$.
The integral over $\app$ in Eq.~\eqref{calc_Gamma_pwl} can then be split
into two regions, ${ \app \leq a }$ and ${ \app \geq a }$. 
For the first region, we write
\begin{align}
\!\! \int_{0}^{a} \!\! \rd \app \, \frac{n_{a} (\app)}{\aout^{2}} \, s_{\ell}^{2} \big[ \tfrac{\ain}{\aout} , \ein , \eout \big] & \, = \!\! \int_{0}^{a} \!\! \rd \app \, \frac{n_{a} (\app)}{a^{2}} \, s_{\ell}^{2} \big[ \tfrac{\app}{a} , \epp , e \big]
\nonumber
\\
& \, = \!\! \int_{0}^{1} \!\! \rd \alpha \, \frac{n_{a} (a \alpha)}{a} \, s_{\ell}^{2} \big[ \alpha , \epp , e \big]
\nonumber
\\
& \, = \frac{N_{0}}{a_{0}^{2}} \bigg( \frac{a}{a_{0}} \bigg)^{1 - \gamma} \!\! \int_{0}^{1} \!\! \rd \alpha \, \alpha^{2 - \gamma} \, s_{\ell}^{2} \big[ \alpha , \epp , e \big] .
\label{calc_below_Gamma_pwl}
\end{align}
and a very similar calculation can be carried out for the second region to get
\begin{equation}
\!\! \int_{a}^{+ \infty} \!\! \rd \app \, \frac{n_{a} (\app)}{\aout^{2}} \, s_{\ell}^{2} \big[ \tfrac{\ain}{\aout} , \ein , \eout \big] = \frac{N_{0}}{a_{0}^{2}} \, \bigg( \frac{a}{a_{0}} \bigg)^{1 - \gamma} \!\! \int_{0}^{1} \!\! \rd \alpha \, \alpha^{\gamma - 2} \, s_{\ell}^{2} \big[ \alpha , e , \epp \big] .
\label{calc_above_Gamma_pwl}
\end{equation}
In order to shorten the notations, let us introduce the dimensionless integrals
\begin{align}
I_{\ell}^{(1)} [ p , e ] & \, = \!\! \int \!\! \rd \epp \, f_{e} (\epp) \!\! \int_{0}^{1} \!\! \rd \alpha \, \alpha^{p} \, s_{\ell}^{2} \big[ \alpha , e , \epp \big] ,
\nonumber
\\
J_{\ell}^{(1)} [ p , e ] & \, = \!\! \int \!\! \rd \epp \, f_{e} (\epp) \!\! \int_{0}^{1} \!\! \rd \alpha \, \alpha^{p} \, s_{\ell}^{2} \big[ \alpha , \epp , e \big] ,
\label{def_I1_J1_pwl}
\end{align}
where one should pay attention to the order of the arguments of ${ s_{\ell}^{2} }$.
This allows us then to rewrite Eq.~\eqref{calc_Gamma_pwl} as
\begin{equation}
\Gamma^{2} (\bK) = \Gamma_{0}^{2} \, \frac{f_{\Gamma^{2}} (e)}{1 - e^{2}} \, \bigg( \frac{a}{a_{0}} \bigg)^{- \gamma} ,
\label{short_Gamma_pwl}
\end{equation}
where we introduced the amplitude $\Gamma_{0}$ and the dimensionless function ${ f_{\Gamma^{2}} (e) }$ as
\begin{align}
\Gamma_{0}^{2} & \, = \frac{1}{4 \pi} \, \frac{G N_{0} \avmsq}{\mBH a_{0}^{3}} ,
\nonumber
\\
f_{\Gamma^{2}} (e) & \, = \sum_{\ell} B_{\ell} \bigg\{ I_{\ell}^{(1)} [\gamma - 2 , e] + J_{\ell}^{(1)} [2 - \gamma , e] \bigg\} .
\label{def_Gamma0_fGamma_pwl}
\end{align}
In Fig.~\ref{fig:PowerLawf}, we illustrate the dependence of ${ f_{\Gamma^{2}} (e) }$, assuming a thermal eccentricity distribution, ${ f_{e} (e) = 2 e }$, and different cusp's profiles.

Let us now pursue a similar approach to compute the coherence time, $\Tc^{\rt}$,
as introduced in Eq.~\eqref{eq:Tct_def}.
When expanding the r.h.s.\ of that equation, one gets
\begin{equation}
\Gamma^{2} (\bK) \, \Tc^{\rt} (\bK) = \Gamma_{0} \, \frac{a_{0}^{2}}{a (1 - e^{2})} \sum_{\ell} \frac{B_{\ell}}{\sqrt{A_{\ell} /2}} \!\! \int \rd \epp \, \frac{\sqrt{1 - \eppsq} \, f_{e} (\epp)}{\sqrt{f_{\Gamma^{2}} (\epp)}} \!\! \int \!\! \rd \app \, \frac{1}{\aout^{2}} \, \bigg( \frac{\app}{a_{0}} \bigg)^{\tfrac{4 - \gamma}{2}} s_{\ell}^{2} \big[ \tfrac{\ain}{\aout} , \ein , \eout \big] . 
\label{calc_Tc_pwl}
\end{equation}
The integral over ${ \rd \app }$ can be carried out following the same approach as in Eqs.~\eqref{calc_below_Gamma_pwl} and~\eqref{calc_above_Gamma_pwl} to get
\begin{align}
\!\! \int_{0}^{a} \!\! \rd \app \, \frac{1}{\aout^{2}} \bigg( \frac{\app}{a_{0}} \bigg)^{\tfrac{4 - \gamma}{2}} s_{\ell}^{2} \big[ \tfrac{\ain}{\aout} , \ein , \eout \big] & \, = \frac{1}{a_{0}} \, \bigg( \frac{a}{a_{0}} \bigg)^{\tfrac{2 - \gamma}{2}} \!\! \int_{0}^{1} \!\! \rd \alpha \, \alpha^{\tfrac{4 - \gamma}{2}} \, s_{\ell}^{2} \big[ \alpha , \epp , e \big] ,
\nonumber
\\
\!\! \int_{a}^{+ \infty} \!\! \rd \app \, \frac{1}{\aout^{2}} \bigg( \frac{\app}{a_{0}} \bigg)^{\tfrac{4 - \gamma}{2}} s_{\ell}^{2} \big[ \tfrac{\ain}{\aout} , \ein , \eout \big] & \, = \frac{1}{a_{0}} \, \bigg( \frac{a}{a_{0}} \bigg)^{\tfrac{2 - \gamma}{2}} \!\! \int_{0}^{1} \!\! \rd \alpha \, \alpha^{\tfrac{\gamma - 4}{2}} \, s_{\ell}^{2} \big[ \alpha , e , \epp \big] .
\label{calc_below_above_Tc_pwl}
\end{align}
Similarly to Eq.~\eqref{def_I1_J1_pwl}, in order to shorten the notations, we define the dimensionless integrals
\begin{align}
I_{\ell}^{(2)} [p , e] & \, = \!\! \int \!\! \rd \epp \, \frac{\sqrt{1 - \eppsq} \, f_{e} (\epp)}{\sqrt{f_{\Gamma^{2}} (\epp)}} \!\! \int_{0}^{1} \!\! \rd \alpha \, \alpha^{p} \, s_{\ell}^{2} \big[ \alpha , e , \epp \big] ,
\nonumber
\\
J_{\ell}^{(2)} [p , e] & \, = \!\! \int \!\! \rd \epp \, \frac{\sqrt{1 - \eppsq} \, f_{e} (\epp)}{\sqrt{f_{\Gamma^{2}} (\epp)}} \!\! \int_{0}^{1} \!\! \rd \alpha \, \alpha^{p} \, s_{\ell}^{2} \big[ \alpha , \epp , e \big] ,
\label{def_I2_J2_pwl}
\end{align}
where once again, one should pay attention to the order of the arguments of $s_{\ell}^{2}$.
Gathering all these elements, Eq.~\eqref{calc_Tc_pwl} gives us
the needed expression of $\Tc^{\rt}$.
It reads
\begin{equation}
\Tc^{\rt} (\bK) = T_{0} \, f_{T} (e) \, \bigg( \frac{a}{a_{0}} \bigg)^{\gamma/2} .
\label{short_Tct_pwl}
\end{equation}
where we introduced the amplitude $T_{0}$ and the dimensionless function ${ f_{T} (e) }$ as
\begin{equation}
T_{0} = \frac{1}{\Gamma_{0}} ; \;\;\; f_{T} (e) = \frac{1}{f_{\Gamma^{2}} (e)} \sum_{\ell} \frac{B_{\ell}}{\sqrt{A_{\ell}/2}} \bigg\{ I_{\ell}^{(2)} \big[ \tfrac{\gamma - 4}{2} , e \big] + J_{\ell}^{(2)} \big[ \tfrac{4 - \gamma}{2} , e \big] \bigg\} .
\label{def_T0_fT_pwl}
\end{equation}
In Fig.~\ref{fig:PowerLawf}, for a thermal eccentricity distribution, we illustrate how one can assume ${ f_{\Gamma^{2}}^{1/2} (e) \, f_{T} (e) \simeq 0.4 }$ independent of $e$ and the considered cups's power index $\gamma$.

\end{document}